\documentclass[letterpaper,10pt,aps,pre,notitlepage,reprint,superscriptaddress,floatfix]{revtex4-1}
\usepackage{bm,amssymb,amsmath,amsfonts,graphicx}
\usepackage{epstopdf}

\begin{document}

\title{Inferring interaction partners from protein sequences}
\author{Anne-Florence Bitbol}
\email{anne-florence.bitbol@upmc.fr}
\affiliation{Lewis-Sigler Institute for Integrative Genomics, Princeton University, Princeton, NJ 08544, USA}
\affiliation{Department of Physics, Princeton University, Princeton, NJ 08544, USA}
\affiliation{Sorbonne Universit\'es, Universit\'e Pierre et Marie Curie - Paris 6, CNRS, Laboratoire Jean Perrin (UMR 8237), F-75005, Paris, France}
\author{Robert S. Dwyer}
\affiliation{Department of Molecular Biology, Princeton University, Princeton, NJ 08544, USA}
\author{Lucy J. Colwell} 
\email{ljc37@cam.ac.uk}
\altaffiliation{L.J.C. and N.S.W. contributed equally to this work.}
\affiliation{Department of Chemistry, University of Cambridge, Lensfield Road, Cambridge CB2 1EW, United Kingdom}
\author{Ned S. Wingreen}
\email{wingreen@princeton.edu}
\affiliation{Lewis-Sigler Institute for Integrative Genomics, Princeton University, Princeton, NJ 08544, USA}
\affiliation{Department of Molecular Biology, Princeton University, Princeton, NJ 08544, USA}

\begin{abstract}
Specific protein-protein interactions are crucial in the cell, both to ensure the formation and stability of multi-protein complexes, and to enable signal transduction in various pathways. Functional interactions between proteins result in coevolution between the interaction partners, causing their sequences to be correlated. Here we exploit these correlations to accurately identify which proteins are specific interaction partners from sequence data alone. Our general approach, which employs a pairwise maximum entropy model to infer couplings between residues, has been successfully used to predict the three-dimensional structures of proteins from sequences. Thus inspired, we introduce an iterative algorithm to predict specific interaction partners from two protein families whose members are known to interact. We first assess the algorithm's performance on histidine kinases and response regulators from bacterial two-component signaling systems. We obtain a striking 0.93 true positive fraction on our complete dataset without any \textit{a priori} knowledge of interaction partners, and we uncover the origin of this success. We then apply the algorithm to proteins from ATP-binding cassette (ABC) transporter complexes, and obtain accurate predictions in these systems as well. Finally, we present two metrics that accurately distinguish interacting protein families from non-interacting ones, using only sequence data.
\end{abstract}

\maketitle

\section*{Significance}
Specific protein-protein interactions play crucial roles in the stability of multi-protein complexes and in signal transduction. Thus, mapping these interactions is key to a systems-level understanding of cells. Systematic experimental identification of protein interaction partners is still challenging. However, a large and rapidly growing amount of sequence data is now available. Is it possible to identify which proteins interact just from their sequences? We propose an approach based on sequence covariation, building on methods used with success to predict the three-dimensional structures of proteins from sequences alone. Our method identifies specific interaction partners with high accuracy among the members of several ubiquitous prokaryotic protein families, and provides a way to predict protein-protein interactions directly from sequence data.

\section*{Introduction}
Many key cellular processes are carried out by interacting proteins. For instance, specific protein-protein interactions ensure proper signal transduction in various pathways. Hence, mapping specific protein-protein interactions is central to a systems-level understanding of cells, and has broad applications to areas such as drug targeting. High-throughput experiments have recently elucidated a substantial fraction of protein-protein interactions in a few model organisms~\cite{Rajagopala14}, but such experiments remain challenging. Meanwhile, there has been an explosion of available sequence data. Can we exploit this abundant new sequence data to identify specific protein-protein interaction partners?

Specific interactions between proteins imply evolutionary constraints on the interacting partners. For instance, mutation of a contact residue in one partner generally impairs binding, but may be compensated by a complementary mutation in the other partner. This co-evolution of interaction partners results in correlations between their amino-acid sequences. Similar correlations exist within single proteins, for example between amino acids that are in contact in the folded protein. However, the simple fact of a correlation between residues in a multiple sequence alignment is only weakly predictive of a three-dimensional contact~\cite{Altschuh87,Lockless99,Skerker08}, as correlation can also stem from indirect effects. Fortunately, global statistical models allow direct and indirect interactions to be disentangled~\cite{Lapedes99,Burger08,Weigt09}. In particular, the maximum entropy principle~\cite{Jaynes57} specifies the least-structured global statistical model consistent with the one- and two-point statistics of an alignment~\cite{Lapedes99}. This approach has recently been used with success to determine three-dimensional protein structures from sequences~\cite{Marks11,Sulkowska12,Jones12}, to predict mutational effects~\cite{Dwyer13,Cheng14,Figliuzzi16}, and to find residue contacts between known interaction partners~\cite{Weigt09,Procaccini11,Baldassi14,Ovchinnikov14,Hopf14,Feinauer16}. Pairwise maximum entropy models have also been used productively in various other fields~\cite{Schneidman06,Lezon06,Mora10,Bialek12,Wood12,Ferguson13,Mann14}. 

\begin{figure*}[h t b] 
\centering
\includegraphics[width=.9\linewidth]{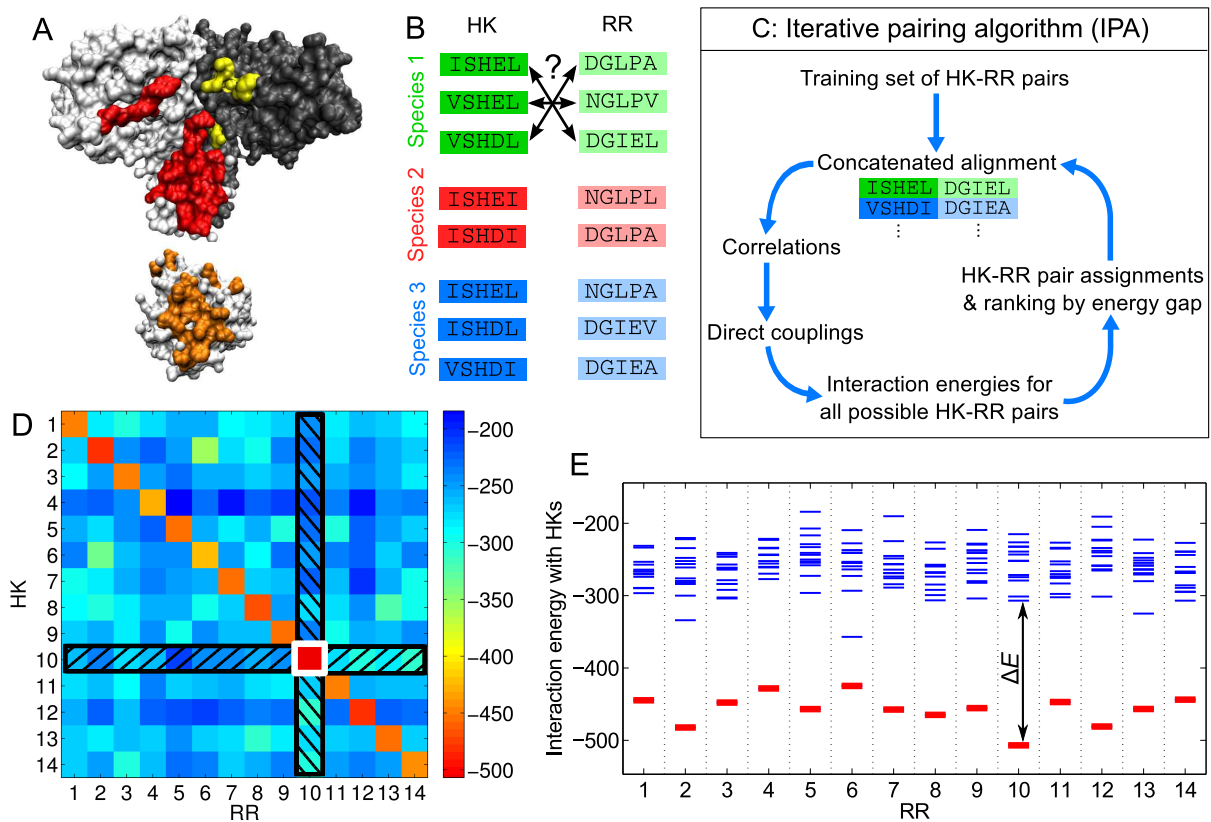}
\caption{Iterative pairing algorithm (IPA). (A) Surface representations of a histidine kinase dimer (HK, top) and a response regulator (RR, bottom), from a co-crystal structure~\cite{Casino09}; the HK-RR contacts in each molecule are highlighted in color. (B) To correctly pair HKs and RRs in each species from their sequences alone, we start from multiple sequence alignments of HKs and RRs, including 64 amino acids from the HK and 112 from the RR. (C) Schematic of the main steps of the IPA. (D, E) Example of HK-RR pair assignment and ranking by energy gap for one species. (D) Color map of the matrix of HK-RR interaction energies in \textit{E. coli} K-12 MG1655 from the final iteration of the IPA performed on our standard dataset, with a training set of $N_{\mathrm{start}}=100$ HK-RR pairs, and an increment step of $N_{\mathrm{increment}}=200$ pairs. As in every IPA iteration and every species, the pair with the lowest interaction energy is selected first (here, HK 10 - RR 10, boxed in white), and this HK and RR are removed from further consideration (black hatches). Then, the next pair with the lowest energy is chosen, and the process is repeated until all HKs and RRs are paired. (E) Energy spectrum from (D), showing the interaction energies with all the HKs for each RR, with the correct HK-RR pairs shown in red. The energy gap $\Delta E$ is shown for RR 10. A confidence score based on the energy gap is used to rank all assigned HK-RR pairs, and this ranking is exploited in order to build the concatenated alignment for the subsequent IPA iteration. See Materials and Methods for details.}
\label{Fig1}
\end{figure*}

Here we present a pairwise maximum entropy approach that uses sequence data to predict interaction partners among the paralogous genes belonging to two interacting protein families. We use histidine kinases (HKs) and response regulators (RRs) from prokaryotic two-component signaling systems (Fig.~\ref{Fig1}A) as our main benchmark. Two-component systems constitute a major class of pathways that enable bacteria to sense and respond to environment signals. Typically, a transmembrane HK senses a signal, autophosphorylates, and transfers its phosphate group to its cognate RR, which induces a cellular response~\cite{Laub07}. Most HKs are encoded in operons together with their cognate RR, so interaction partners are known, which enables us to assess performance. There are often dozens of paralogs of HKs and RRs in each genome, making prediction of interaction partners from sequences alone highly nontrivial. 

To address this challenge, we developed an iterative pairing algorithm (IPA, Fig.~\ref{Fig1}) that pairs proteins based on their effective interaction energies as predicted by a pairwise maximum entropy model. At each iteration, the highest-scoring predicted HK-RR pairs are incorporated into the concatenated sequence alignment from which the model is built. This yields a major increase of predictive accuracy through progressive training of the model. First, we consider the case where the IPA starts with a training set of known HK-RR partners. We obtain good performance even with few training pairs. Then, we show that the IPA can make accurate predictions \emph{starting without any known pairings}, as would be needed to predict novel protein-protein interactions. We trace the origin of this success to the preferential recruitment of new correct pairs by those already in the concatenated alignment. We also demonstrate how multiple random initializations can be leveraged to improve performance. We show that our algorithm works more generally by successfully applying it to ATP-binding cassette (ABC) transporter proteins. Finally, we develop two IPA-based methods that distinguish interacting protein families from non-interacting ones, using only sequence data.

\section*{Results}

\subsection*{Iterative pairing algorithm (IPA)}
We have developed an iterative method to predict interaction partners among the paralogs of two protein families in each species, using just their sequences (Fig.~\ref{Fig1}A,B). In each iteration (Fig.~\ref{Fig1}C; Materials and Methods), we compute correlations between residues from a concatenated alignment (CA) of paired sequences. The initial CA is either built from a training set of correct protein pairs, or made from random pairs, assuming no prior knowledge of interacting pairs. We then infer couplings for all residue pairs using a pairwise maximum entropy model built from the CA using a mean-field approximation~\cite{Marks11,Morcos11}. We calculate the interaction energy for every possible protein pair within each species, by summing the inter-protein couplings assigned by the model. Such ``energies'' capture evolutionary correlations, and correlate to physical energies for lattice proteins~\cite{Jacquin16}. Using these interaction energies, we predict protein pairs (assuming one-to-one specific HK-RR interactions~\cite{Laub07}, Fig.~\ref{Fig1}D). We attribute a confidence score to each predicted protein pair, using the energy gap between this pair and the next best alternative (Fig.~\ref{Fig1}E). The CA is then updated by including the highest-scoring protein pairs, and the next iteration can begin. At each iteration, all pairs in the CA are re-selected based on confidence scores (except initial training pairs, if any), allowing for error correction. 

Unless otherwise specified in what follows, our results were obtained on a standard dataset comprising 5064 HK-RR pairs for which the correct pairings are known from gene adjacency. Each species has on average $\langle m_p\rangle=11.0$ pairs, and at least two pairs (see Materials and Methods).

\subsection*{Starting from known pairings}
We begin by predicting interaction partners starting from a training set of known pairs. To implement this, we pick a random set of $N_\mathrm{start}$ known HK-RR pairs from our standard dataset, and the first IPA iteration uses this concatenated alignment (CA)  to train the model. We blind the pairings of the remaining test set, and predict them. At each subsequent iteration $n>1$, the CA used to retrain the model contains the initial training pairs plus the $(n-1) N_\mathrm{increment}$ highest-scoring predicted pairs from the previous iteration (see Materials and Methods). 

\begin{figure}[h]
\centering
\includegraphics[width=\linewidth]{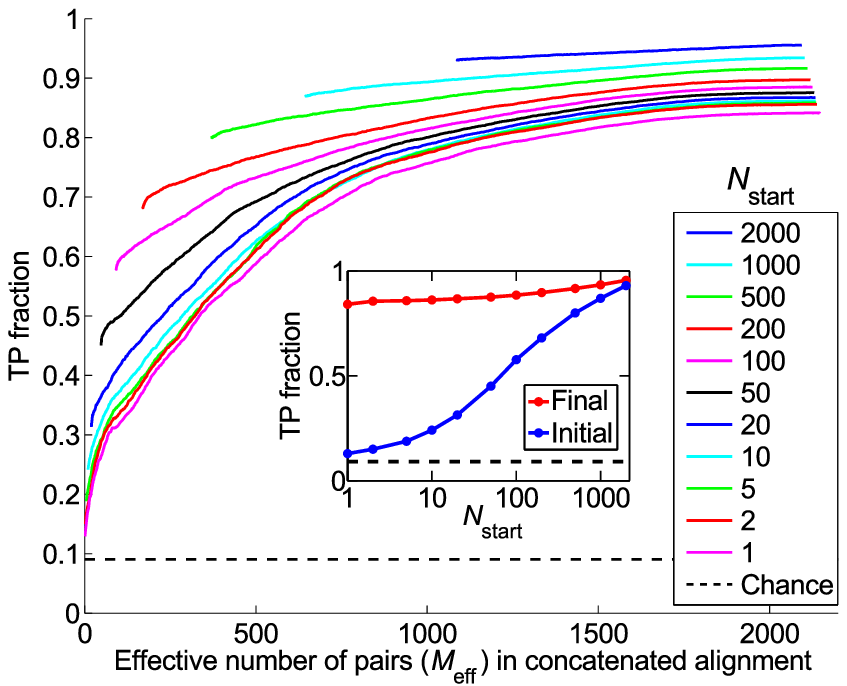}
\caption{Fraction of predicted pairs that are true positives (TP fraction), for different training set sizes $N_{\mathrm{start}}$. Main panel: Progression of the TP fraction during iterations of the IPA. The TP fraction is plotted versus the effective number of HK-RR pairs ($M_{\mathrm{eff}}$, see Supporting Information, Eq.~S1) in the concatenated alignment, which includes $N_{\mathrm{increment}}=6$ additional pairs at each iteration. The IPA is performed on the standard dataset, and all results are averaged over 50 replicates that differ by the random choice of training pairs. Dashed line: Average TP fraction obtained for random HK-RR pairings. Inset: Initial and final TP fractions (at first and last iteration) versus $N_{\mathrm{start}}$.}
\label{Fig2}
\end{figure}

At the first iteration, the fraction of accurately predicted HK-RR pairs (true positive (TP) fraction) depends strongly on $N_\mathrm{start}$, and is close to the random expectation (0.09) for small training sets, ranging from 0.13 at $N_\mathrm{start}=1$ to 0.93 for $N_\mathrm{start}=2000$ (Fig.~\ref{Fig2}, inset, blue curve). The TP fraction increases with subsequent iterations (Fig.~\ref{Fig2}, main panel). Strikingly, the final TP fraction depends only weakly on $N_\mathrm{start}$. For $N_\mathrm{start}=1$, the IPA achieves a final TP fraction of 0.84, a huge increase from the initial value of 0.13 (Fig.~\ref{Fig2}, inset, red curve). This demonstrates the power of iterating: incorporating high-scoring predicted pairs progressively increases the predictive accuracy of the model.

\subsection*{Starting without known pairings} 
Given the success of the IPA with very small training sets, we next ask whether predictions can be made without any prior knowledge of interacting pairs. To test this, we randomly pair each HK with an RR from the same species, and use these 5064 random pairs to train the initial model. At each subsequent iteration $n > 1$, the CA is built just from the $(n-1)N_\mathrm{increment}$ highest-scoring pairs from the previous iteration (see Supporting Information).

\begin{figure}[h]
\centering
\includegraphics[width=\linewidth]{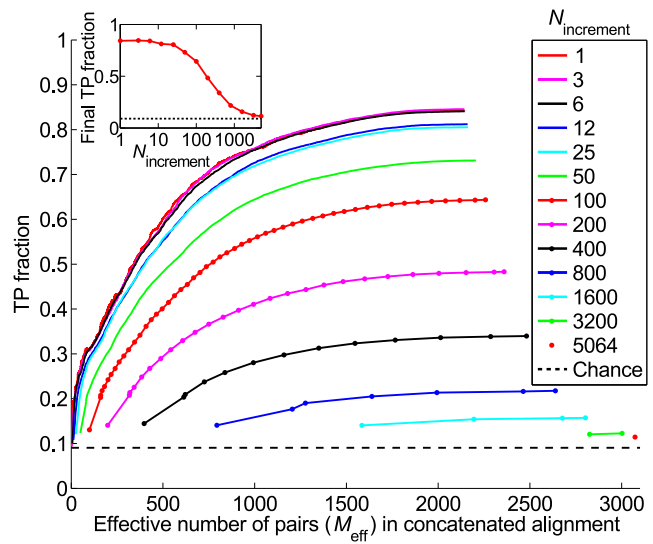}
\caption{Starting from random pairings, i.e. without known pairings. Main panel: TP fraction during iterations of the IPA versus the effective number of HK-RR pairs ($M_{\mathrm{eff}}$) in the concatenated alignment, which includes $N_{\mathrm{increment}}$ additional pairs at each iteration. Different curves correspond to different $N_{\mathrm{increment}}$. The IPA is performed on the standard dataset, and all results are averaged over 50 replicates that differ in their initial random pairings. Note that the first point of each curve corresponds to the second iteration. Dashed line: Average TP fraction obtained for random HK-RR pairings. Inset: Final TP fraction versus $N_{\mathrm{increment}}$.}
\label{Fig3}
\end{figure}

Fig.~\ref{Fig3} shows the progression of the TP fraction for different values of $N_\mathrm{increment}$. It increases in all cases, and the iterative method performs best for small increment steps (Fig.~\ref{Fig3}, inset). The low--$N_\mathrm{increment}$ limit of the final TP fraction is 0.84, identical to that obtained with a single training pair (Fig.~\ref{Fig2}). This striking TP fraction of 0.84 is attained without any prior knowledge of HK-RR interactions: the IPA bootstraps its way toward high predictivity. The low--$N_\mathrm{increment}$ limit is almost reached for $N_\mathrm{increment}= 6$; thus we generically use $N_\mathrm{increment}=6$ to reduce computational time while retaining near-optimal performance. The final TP fraction is robust with respect to different initializations: for $N_\mathrm{increment}=6$, its standard deviation (over 500 replicates) is 0.04.  For the complete dataset (23,424 HK-RR pairs), the IPA yields a TP fraction of 0.93 with no training data (see below).

\subsection*{Training process}
\begin{figure}[h t b]
\centering
\includegraphics[width=\linewidth]{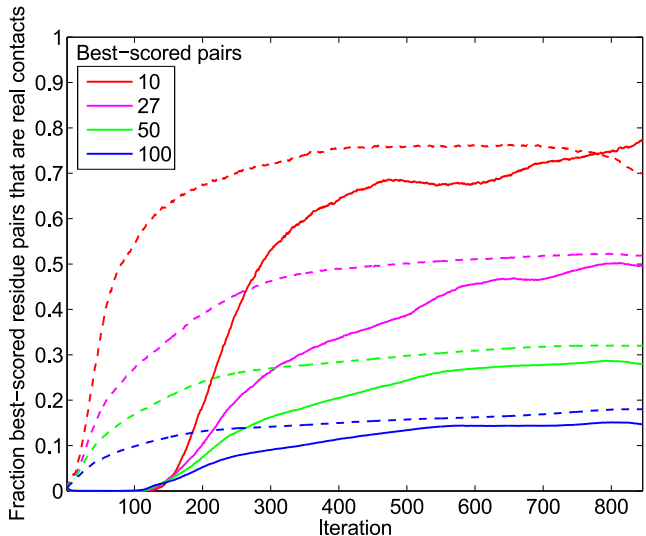}
\caption{Training of the couplings during the IPA. Residue pairs comprised of an HK site and an RR site were scored by the Frobenius norm (i.e. the square root of the summed squares) of the couplings involving all possible residue types at these two sites. The best-scored residue pairs were compared to the 27 HK-RR contacts found experimentally in Ref.~\cite{Casino09}. Solid curves: Fraction of residue pairs that are real contacts (among the $k$ best-scored pairs for four different values of $k$) versus the iteration number in the IPA. Dashed curves: Ideal case, where at each iteration $N_\mathrm{increment}$ randomly-selected correct HK-RR pairs are added to the CA. The overall fraction of residue pairs that are real HK-RR contacts, yielding the chance expectation, is only $3.8\times10^{-3}$. The IPA is performed on the standard dataset with $N_{\mathrm{increment}}=6$, and all data is averaged over 500 replicates that differ in their initial random pairings.}
\label{Fig4}
\end{figure}

The ability to accurately predict interaction partners without training data is surprising. To understand it, we examine the evolution of the model over iterations of the IPA. In a well-trained model, the residue pairs with the largest couplings have been shown to correspond to contacts in the protein complex~\cite{Weigt09,Ekeberg13,Baldassi14}. Up to iteration ${\sim}100-150$ (with $N_\mathrm{increment}= 6$), models starting from random pairings do no better than chance at identifying contacts. Subsequently, they improve rapidly and soon predict contacts nearly as well as models constructed from correct pairs (Figs.~\ref{Fig4} and~S1A). 

At early stages, the model, constructed from few almost random HK-RR pairs, poorly predicts real contacts and correct HK-RR pairs. However, couplings associated to residue pairs that occur in the CA increase, raising the scores of HK-RR pairs with high sequence similarity to those already in the CA, and making them more likely to be added to the CA. We thus examine sequence similarity between the HK-RR pairs in the CA in consecutive iterations. Specifically, we consider two HK-RR pairs to be ``neighbors'' if the sequence identity between the two HKs and between the two RRs are both $>70\%$. We find that sequence similarity is crucial in the early recruitment of new HK-RR pairs to the CA (Fig.~S1B). 

Understanding the initial increase of the TP fraction requires a further observation. In our standard dataset, among all possible, within-species HK-RR pairs, the average number of neighbor pairs per correct HK-RR pair is 9.66, of which 99\% are correct. In contrast, the average number of neighbor pairs per incorrect HK-RR pair is 5.25, of which less than 1\% are correct. Thus, correct pairs are more similar to each other than they are to incorrect pairs, or than incorrect pairs are to each other. We call this the \textit{Anna Karenina effect}, in reference to the first sentence of Tolstoy's novel~\cite{AK}: “All happy families are alike; each unhappy family is unhappy in its own way.” Biologically, this makes sense: each HK-RR pair is an evolutionary unit, so a correct pair is likely to have orthologs of both the HK and the RR in multiple other species, whereas an incorrect pair is less likely to  have orthologs of both the HK and the (non-cognate) RR in other species.  Hence, in early iterations, the number of neighbors recruited per correct pair is higher than per wrong pair (Fig.~S1B), increasing the TP fraction in the CA. To summarize, sequence similarity is crucial at early stages, and the Anna Karenina effect helps to increase the TP fraction in the CA, thus promoting training of the model. This suggests that the IPA might be further enhanced by initially scoring HK-RR pairs based on similarity~\cite{Bradde10}.

\subsection*{Application of the IPA to ABC transporters}
\begin{figure}[h]
\centering
\includegraphics[width=\linewidth]{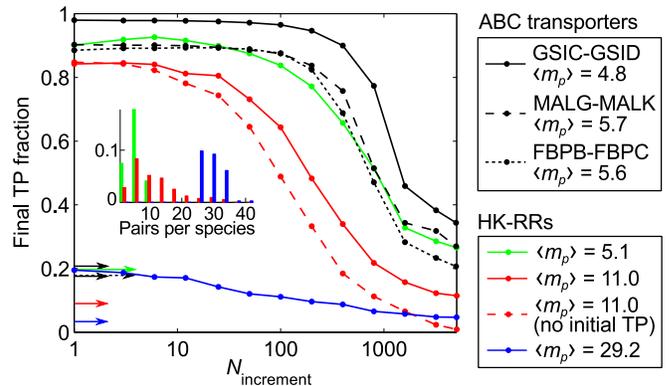}
\caption{Results for ABC transporter pairs and impact of the number of pairs per species. Main panel: Final TP fraction versus $N_{\mathrm{increment}}$ for three different pairs of protein families involved in ABC transport complexes (black curves), and for three HK-RR datasets with different distributions of the number of pairs per species yielding different means $\langle m_p\rangle$ (colored curves). All datasets include $\sim$5000 protein pairs, and the IPA is started from random pairings, apart from the red dashed curve, where it is started from incorrect random pairings. All results are averaged over 50 replicates that differ in their initial pairings. Arrows with the same line style as each curve indicate the average TP fractions obtained for random pairings in each dataset. Inset: Distribution of the number of pairs per species in the three different HK-RR datasets (red: standard dataset; green and blue: datasets comprised of the species with lowest or highest numbers of pairs in the full HK-RR dataset). }
\label{Fig5}
\end{figure}

 To demonstrate the utility of the IPA beyond HK-RRs, we applied it to several protein families involved in ABC transporter complexes. Bacterial genomes typically contain multiple paralogs of these transporters, involved in the translocation of different substances~\cite{Rees09}. We built alignments of homologs of the \textit{Escherichia coli} interacting protein pairs MALG-MALK, FBPB-FBPC, and GSIC-GSID (see Supporting Information). These protein pairs are respectively involved in maltose, iron and glutathione transport systems. The IPA starting from random pairings yields respective final TP fractions 0.90, 0.89, and 0.98 for these pairs in the low-$N_\textrm{increment}$ limit (Fig.~\ref{Fig5}, black curves). These accurate predictions demonstrate the broad applicability of the IPA beyond HK-RRs.

\subsection*{Dependence on features of the dataset}
To apply the IPA approach more widely, it is important to understand what dataset characteristics enable its success. The number of paralogous pairs per species is likely important, since pairing is more difficult when there are more incorrect possibilities. Indeed, higher TP fractions are obtained in datasets with fewer average pairs per species both across different ABC transporter protein pairs and across HK-RR datasets with different numbers of pairs per species (Fig.~\ref{Fig5}). Moreover, the presence of species with a small number of pairs is crucial (Fig.~S2), though in their absence, the TP fraction can be rescued by a sufficiently large training set (Fig.~S3). Perhaps surprisingly, for small $N_\mathrm{increment}$, the final TP fraction does not depend on how many pairs in the initial CA are correct (Fig.~\ref{Fig5}, red curves, and Fig.~S4). Hence, the importance of species with few pairs does not stem from a more favorable initialization. Rather, protein pairs from species with few pairs tend to obtain higher confidence scores since they have fewer competitors, making them more likely to enter the CA at early stages (Fig.~S5). This bias in favor of species with few pairs combines with the Anna Karenina effect to favor correct pairs early in the learning process.

Since sequence similarity is crucial at early iterations, it should strongly impact performance. Indeed, a lower final TP fraction (0.58 vs. 0.84) is obtained in an HK-RR dataset where no two correct pairs are $>70\%$ identical, but it can be rescued by a sufficiently large training set (Fig.~S6).

Another important parameter is dataset size. For HK-RRs, the final TP fraction increases steeply above ${\sim}1000$ sequences, and saturates above ${\sim}10,000$ (Fig.~S7). For the complete dataset (23,424 HK-RR pairs, see Materials and Methods), we obtain a striking final TP fraction of 0.93. Larger datasets imply closer neighbors, which is favorable to the success of the IPA, particularly in the absence of training data. 

\subsection*{Optimization}
To improve the predictive ability of the IPA, we exploit multiple different random initializations of the CA. For each possible, within-species HK-RR pair, we calculate the fraction $f_r$ of replicates of the IPA in which this pair is predicted. High $f_r$ values are excellent predictors of correct pairs, outperforming average TP fractions from individual replicates (Fig.~S8). The quality of $f_r$ as a classifier is demonstrated by the area under the receiver operating characteristic: it is 0.991, very close to 1 (ideal). The very high TP fraction of the pairs with highest $f_r$ can be exploited by taking some as training pairs and running the IPA again. This ``rebootstrapping'' process can be iterated, yielding further performance increases, particularly for small datasets (Fig.~S9).

\subsection*{Determining whether two protein families interact}
\begin{figure}[h]
\centering
\includegraphics[width=\linewidth]{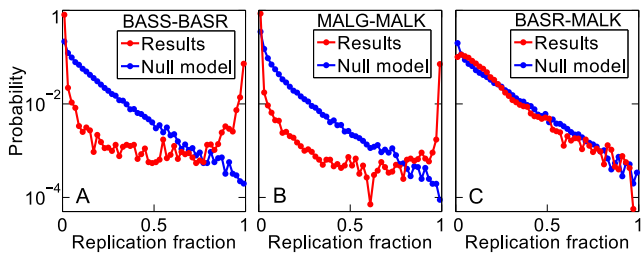}
\caption{An IPA-derived signature of protein-protein interactions. For three pairs of protein families, we compute the fraction $f_r$ of IPA replicates in which each possible within-species protein pair is predicted as a pair. (A and B) Protein families with known interactions: (A) BASS-BASR homologs and (B) MALG-MALK homologs; (C) Protein families with no known interaction (BASR-MALK homologs). Red curves: Distribution of $f_r$ obtained for each alignment. Blue curves: Same distribution obtained by running the IPA on alignments where each column is scrambled (null model). Alignments include $\sim$5000 pairs, with $\left<m_p\right>\approx 5$, and each distribution is estimated from 500 IPA replicates that differ in their initial random pairings, using $N_{\mathrm{increment}}=50$.}
\label{Fig6}
\end{figure}

The IPA correctly predicts most interacting protein pairs no matter which initial random pairing is used. This suggests that the distribution of replication fractions $f_r$ (over all possible within-species pairs) should distinguish pairs of protein families that interact from those that do not. To test this idea, we consider three pairs of families with similar $\left<m_p\right>$: the subset of HK-RRs homologous to BASS-BASR, the homologs of the interacting ABC transporter proteins MALG-MALK, and a pair with no known interaction, homologs of BASR-MALK. For both interacting protein families, the distribution of replication fractions $f_r$ strongly favors values close to 0, mostly corresponding to wrong pairs, and close to 1, mostly corresponding to correct pairs (Fig.~\ref{Fig6}A-B). No such bimodality is observed for BASR-MALK (Fig.~\ref{Fig6}C). We constructed null models for each dataset by randomly scrambling the amino acids at each site (column) of the alignment, thus retaining conservation while removing correlations. For BASR-MALK, the observed $f_r$ distribution is very similar to the null-model distribution, while for both interacting pairs the results and the null strongly differ (Fig.~\ref{Fig6}). The standard HK-RR dataset can be similarly contrasted with an HK-RR dataset lacking correct pairs (Fig.~S10). Comparing the observed $f_r$ distribution to the null thus distinguishes interacting from non-interacting  protein families using sequence data alone. For these family pairs, the difference in $f_r$ distributions is visible down to dataset sizes $M\sim$500. Another signature of interacting families is the strength of the top predicted contacts (Fig.~S11).

\section*{Discussion} 

We have presented a method to infer interaction partners among two protein families with multiple paralogs, using only sequence data. Our approach is based on pairwise maximum entropy models, which have proved successful at predicting residue contacts between known interaction partners~\cite{Weigt09,Procaccini11,Baldassi14,Ovchinnikov14,Hopf14,Feinauer16}. To our knowledge, the important problem of predicting interaction partners among paralogs from sequences has only been addressed by Burger and van Nimwegen ~\cite{Burger08}, who used a Bayesian network method. Pairwise maximum entropy-based approaches were later shown to outperform this method for orphan HK-RR partner predictions, starting from a substantial training set of partners known from gene adjacency~\cite{Procaccini11}. Importantly, our method enables partner prediction without any initial known pairs, whereas even the seminal study~\cite{Burger08} included a training set via species that contain only a single pair. This lack of a training set is important to predict novel protein-protein interactions, since in this context no prior knowledge of interacting pairs would be available.

We first benchmarked our iterative pairing algorithm (IPA) on HK-RR pairs. The top-scoring predicted HK-RR pairs are progressively incorporated into the concatenated alignment used to build the maximum entropy model. This enables progressive training of the model, providing major increases in predictive accuracy. Strikingly, the IPA is very successful even in the absence of any prior knowledge of HK-RR interactions, yielding a 0.93 TP fraction on our complete dataset. The success of the IPA with no training data relies on initial recruitment of pairs by sequence similarity. Correct pairs are more similar to one another than incorrect pairs, favoring recruitment of correct pairs - a process we called the ``Anna Karenina effect''. 

IPA performance is best for large datasets (with strong sequence similarity), and when species with few pairs are included. The first condition is easily met for HK-RRs (a 0.84 TP fraction was obtained with 5064 pairs, out of 23,424, and our rebootstrapping approach yields a 0.64 TP fraction even for a dataset of only 502 pairs, Fig.~S9B) and is realized for a large and growing number of other protein families. Indeed, in the protein family database PfamA-30~\cite{Finn16}, 62\% of the 15,701 entries classified as ``domains'' or ``families'' comprise more than 500 distinct Uniprot sequences. The mean number of paralogs per species in PfamA-30 domains or families is 3.9, so the HK-RR system actually constitutes an unusually difficult case in this respect~\cite{Laub07}. The success we obtained for ABC transporter proteins, which form permanent complexes, while HK-RRs interact transiently, further points to the broad applicability of the IPA. So far we have only applied the IPA to one-to-one interactions, but the method should be fruitful beyond this domain. 

Our approach could be combined with those of Refs.~\cite{Weigt09,Procaccini11,Cheng14,Baldassi14,Ovchinnikov14,Hopf14,Feinauer16} to improve protein complex structure prediction. It solves the major issue~\cite{Ovchinnikov14,Hopf14,Feinauer16} of finding the correct interaction partners among paralogs, which is a prerequisite for accurate contact prediction. In particular, better paralog-partner predictions will help extend accurate contact prediction to currently-inaccessible cases such as eukaryotic proteins, for which genome organization cannot be used to find partners. 

Finally, we have introduced two distinct IPA-based signatures that distinguish between interacting and non-interacting protein families. These results pave the way toward predicting novel protein-protein interactions between protein families from sequence data alone.

\section*{Materials and Methods}
Extended Materials and Methods are presented in the Supporting Information.
\subsection*{HK-RR dataset}
Our dataset was built from the P2CS database~\cite{Barakat09,Ortet15}, which includes two-component system proteins from all fully-sequenced prokaryotic genomes. All data can thus be accessed online. We considered the protein domains through which HKs and RRs interact, namely the Pfam HisKA domain present in most HKs (64 amino acids) and the Pfam Response\_reg domain present in all RRs (112 amino acids). We focused on proteins with known partners, i.e. those encoded in the genome in pairs containing an HK and an adjacent RR. Discarding species with only one pair, for which pairing is unambiguous, we obtained a complete dataset of 23,424 HK-RR pairs from 2102 species. A smaller ``standard dataset'' of 5064 pairs from 459 species was extracted by picking species randomly.

\subsection*{Iterative pairing algorithm (IPA)}

Here, we summarize each of the steps of an iteration of the IPA (Fig.~\ref{Fig1}C).

\textbf{1. Correlations.}
Each iteration begins by the calculation of empirical correlations from the CA of paired HK-RR sequences. The empirical one- and two-site frequencies, $f_i(\alpha)$ and $f_{ij}(\alpha,\beta)$, of occurrence of amino-acid states $\alpha$ (or $\beta$) at each site $i$ (or $j$) are computed for the CA, using a re-weighting of similar sequences, and a pseudocount correction (Eqs.~S1-S4)~\cite{Weigt09,Procaccini11,Marks11,Morcos11}. The correlations are then computed as
\begin{equation}
C_{ij}(\alpha,\beta)=f_{ij}(\alpha,\beta)-f_i(\alpha)f_j(\beta)\,.
\label{Cij_mt}
\end{equation}

\textbf{2. Couplings.}
Next, we construct a pairwise maximum entropy model of the CA (Eq.~S6). It involves one-body fields $h_i$ at each site $i$ and (direct) couplings $e_{ij}$ between all sites $i$ and $j$, which are determined by imposing consistency of the pairwise maximum entropy model with the empirical one- and two-point frequencies of the CA (Eqs.~S7-S8). We use the mean-field approximation~\cite{Morcos11,Marks11}: couplings are obtained by inverting the matrix of correlations:
\begin{equation}
e_{ij}(\alpha,\beta)=C^{-1}_{ij}(\alpha,\beta)\,.
\label{eij}
\end{equation}
We then transform to the zero-sum gauge~\cite{Weigt09,Ekeberg13}.

\textbf{3. Interaction energies for all possible HK-RR pairs.}
The interaction energy $E$ of each possible HK-RR pair within each species is calculated as a sum of couplings:
\begin{equation}
E\left(\alpha_1,...,\alpha_{L_\mathrm{HK}},\alpha_{L_\mathrm{HK}+1},...,\alpha_L\right)=\sum_{i=1}^{L_\mathrm{HK}}\sum_{j=L_\mathrm{HK}+1}^{L}e_{ij}(\alpha_i,\alpha_j)\,,
\end{equation}
where $L_\mathrm{HK}$ denotes the length of the HK sequence and $L$ that of the concatenated HK-RR sequence.

\textbf{4. HK-RR pair assignments and ranking by gap.}
In each species, the pair with the lowest interaction energy is selected first, and the HK and RR involved are removed from further consideration, assuming one-to-one HK-RR matches (Fig.~\ref{Fig1}D). Then, the pair with the next lowest energy is chosen, until all HKs and RRs are paired. Each pair is scored at assignment by a confidence score $\Delta E/(n+1)$, where $\Delta E$ is the energy gap (Fig.~\ref{Fig1}E), and $n$ is the number of lower-energy matches discarded in assignments made previously, within that species and at that iteration (Fig.~S12). All the assigned HK-RR pairs are then ranked in order of decreasing confidence score. 

 \textbf{5. Incrementation of the CA.}
 At each iteration $n>1$, the $(n-1)N_\mathrm{increment}$ assigned pairs that had the highest confidence scores at iteration $n-1$ are included in the CA. In the presence of an initial training set, the $N_\mathrm{start}$ training pairs are also included in the CA. Without a training set, the initial CA is built by randomly pairing each HK of the dataset to an RR from the same species, and for $n>1$, the CA only contains the above-mentioned $(n-1)N_\mathrm{increment}$ assigned pairs. Once the new CA is constructed, the next iteration can start.

\section*{Acknowledgments}
We thank Mohamed Barakat and Philippe Ortet for sharing and discussing specifically-formatted datasets built from the P2CS database. AFB acknowledges support from the Human Frontier Science Program. This research was supported in part by NIH Grant R01-GM082938 (AFB and NSW) and by NSF Grant PHY–1305525 (NSW), Marie Curie Career Integration Grant 631609 (LJC), a Next Generation Fellowship (LJC), and the Eric and Wendy Schmidt Transformative Technology Fund.

\section*{Author contributions}
A.F.B., R.S.D., L.J.C. and N.S.W. designed research, A.F.B., L.J.C. and N.S.W. performed research, analyzed data, and wrote the paper.

\section*{Note}
While submitting this manuscript, we learned that T. Gueudre, C. Baldassi, M. Zamparo, M. Weigt, and A. Pagnani are preparing a related paper on predicting interacting paralog pairs.


\clearpage

\newpage

\centerline{\Large{\textbf{\textsf{SUPPORTING INFORMATION}}}}
\vspace{0.5cm}
\normalsize
\renewcommand{\thefigure}{S\arabic{figure}}
\setcounter{figure}{0} 
\renewcommand{\theequation}{S\arabic{equation}}
\setcounter{equation}{0} 
\renewcommand{\thetable}{S\arabic{table}}
\setcounter{table}{0} 

\centerline{\textbf{\textsf{EXTENDED MATERIALS AND METHODS}}}

\section{Dataset construction}
\subsection{Complete HK-RR dataset}
Our dataset is built using the online database P2CS (\texttt{http://www.p2cs.org/})~\cite{Barakat09,Ortet15}, which includes two-component-system proteins from all fully-sequenced prokaryotic genomes. In the construction of P2CS, these proteins were identified by searching genomes for two-component system domains from the Pfam (\texttt{http://pfam.xfam.org/}) and SMART (\texttt{http://smart.embl-heidelberg.de/}) libraries. We kept only chromosome-encoded proteins, due to strong variability in plasmid presence. We also excluded hybrid and unorthodox proteins, which involve both HK and RR domains in the same protein, since the energetics of partnering is different and often less constraining for such proteins~\cite{Cheng14}. In HKs, there are different domain variants in the vicinity of the N-terminal Histidine-containing phosphoacceptor site, including the region that interacts with RRs. These variants are classified into several different Pfam domain families, which are all members of the His\_Kinase\_A domain clan (CL0025). In order to reliably align all HK sequences, we chose to focus on only one of these Pfam domain families, HisKA (PF00512). Proteins containing a HisKA domain account for the majority (64\%) of all chromosome-encoded, non-hybrid, orthodox HKs in P2CS.  

Proteins in P2CS are annotated based on genetic organization~\cite{Ortet15}. As our aim was to benchmark our method on known, specific interaction partners, we only considered HKs and RRs that are encoded by adjacent genes. Note that 67\% of all chromosome-encoded, non-hybrid, orthodox HKs in P2CS are from such pairs. Suppressing the (rare) HKs with multiple HisKA domains and RRs with multiple Response\_reg domains for which the pairing of domains is ambiguous, this yields 23,632 distinct pairs that differ in either sequence or species. Discarding the 208 pairs from species with only one such pair (see discussion below) yields a dataset of 23,424 HK-RR pairs. Grouping together sequences with mean Hamming distance per site $<0.3$ (i.e. with 70\% sequence identity or more) to estimate sequence diversity yields an effective number of HK-RR pairs $M_\mathrm{eff}=5391$ in the complete dataset.

These 23,424 HK-RR pairs are from 2102 different species, with numbers of pairs per species ranging from 2 to 41, with mean $\langle m_p\rangle=11.1$. The distribution of the number of pairs per species in our complete dataset is shown in Fig.~\ref{FigS_SeqSim}A.

\subsection{Standard HK-RR dataset}
In most of our work, we focused on a smaller ``standard dataset'' extracted from this complete dataset, both because protein families that possess as many members as the HKs and RRs are atypical, and in view of computational time constraints. Note, however, that our IPA was used to make predictions on the complete dataset, yielding a striking 0.93 final TP fraction (Fig.~\ref{FigS_NumberSeqs}). 

Our standard dataset was constructed by picking species randomly.  Once 43 species with one single pair are suppressed (see discussion below), it comprises 5064 pairs from 459 species, with an average number of pairs per species $\langle m_p\rangle=11.0$, which is very close to that of the complete dataset (see Fig.~\ref{FigS_SeqSim}A for the distributions of the number of pairs per species). Grouping together sequences with mean Hamming distance per site $<0.3$ to estimate sequence diversity yields an effective number of HK-RR pairs $M_\mathrm{eff}=2091$ in the standard dataset.

\subsection{Suppressing species with a single pair}
In our datasets, we discarded sequences from species that contain only one known pair, for which pairing is therefore unambiguous. This allowed us to quantitatively assess the impact of training set size ($N_\mathrm{start}$) without the inclusion of an implicit training set via these pairs. More importantly, this enabled us to address prediction in the absence of any known pairs (no training set), which is crucial for predicting unknown protein-protein interactions between protein families, since no training set is then available. For other purposes, pairs from species with only one known pair might be included as a training set (but then one would need to be sure that they are actually interacting,
because any error in the training set would be detrimental for the model). 
In our standard HK-RR dataset, if the 43 pairs from species with a single pair are treated as a training set instead of being discarded, the IPA yields a final TP fraction of 0.88 (vs. 0.84 starting from random pairings, i.e. in the absence of any training set). This value is the same as the one obtained for $N_\mathrm{start}=50$ (0.88, value averaged over 50 different random choices of the 50 training pairs, see Fig.~\ref{Fig2}). Interestingly, by exploiting multiple random initializations, a TP fraction of 0.89 is reached starting from random pairings (Fig.~\ref{FigS_RepFracRank}).

\subsection{Multiple sequence alignment of HKs and RRs}
All HKs in our dataset were aligned to the profile hidden Markov model (HMM) representing the Pfam HisKA domain (PF00512) using the \texttt{hmmalign} tool from the HMMER suite (\texttt{http://hmmer.org/}). Similarly, all RRs were aligned to the profile HMM representing the Pfam Response\_reg domain (PF00072). The aligned sequences of each HK were then concatenated to those of their RR partner, yielding a concatenated multiple sequence alignment. The length of each concatenated sequence is $L=176$ amino acids, among which the $L_\mathrm{HK}=64$ first amino acids are from the HK, and the remaining 112 amino acids are from the RR. The full length of these sequences was kept throughout. 

\subsection{Dataset construction for ABC transporter proteins}
While we used HK-RRs as the main benchmark for the IPA, we also applied it to several pairs of protein families involved in ABC (ATP-binding cassette) transporter complexes. These ubiquitous complexes enable ATP-powered translocation of various substances through membranes~\cite{Rees09}. As in the case of HK-RRs, bacterial genomes typically contain multiple paralogs of these transporters, and actual pairings are known from genome proximity, enabling us to assess the success of the IPA. 

We built paired alignments of homologs of the \textit{Escherichia coli} interacting protein pairs MALG-MALK, FBPB-FBPC, and GSIC-GSID, all involved in ABC transporter complexes, using a method adapted from Ref.~\cite{Ovchinnikov14} and \texttt{http://gremlin.bakerlab.org/}. First, the homologs of each protein were retrieved from Uniprot (\texttt{http://www.uniprot.org/}) using \texttt{hhblits} from the HH-suite (\texttt{https://github.com/soedinglab/hh-suite}) with main options \texttt{-n 8 -e 1E-20}. Then \texttt{hhfilter} from the HH-suite was run with options \texttt{-id 100 -cov 75} to only retain the homologs that have at least 75\% coverage. In order to focus on the relevant conserved domains involved in binding, as we did for HK-RRs, we then used \texttt{hmmsearch} from the HMMER suite to align a subsequence of each homolog to the profile HMM of the appropriate domain from Pfam. These domains are ABC\_tran (PF00005) for
MALK, and BPD\_transp\_1 (PF00528) for all other ABC transporter proteins considered here. For each pair of interacting protein families, sequences from the same species (found via the OX/OS field in the Uniprot headers) were then paired to their interacting partner by genome proximity (assessed via the Uniprot accession numbers, and using a maximum allowed difference of 20 between these IDs). These pairings enabled us to evaluate IPA performance (Fig.~\ref{Fig5}), as in the HK-RR case. Note that the paired alignment of HK-RRs homologous to BASS-BASR was constructed in the same way as the alignments of these ABC-transporter protein pairs.

We also considered a pair of protein families with no known interactions: BASR homologs (Response\_reg domain) and MALK homologs (ABC\_tran domain). These two protein families have very different biological functions, and no interaction between BASR and MALK has been reported in the STRING database (\texttt{http://string-db.org/}).

As in the case of HK-RRs, for each pair of protein families, we worked on subsets of $\sim 5000$ protein pairs extracted from the complete dataset by randomly picking species, and we discarded species with a single pair.

\section{Statistics of the concatenated alignment (CA)}

Henceforth, as in the main text, we will present our general method in the specific case of HK-RRs. Note that we applied it in the exact same way to ABC transporter protein pairs.

Let us consider a CA of paired HK-RR sequences. At each site $i\in\{1,..,L\}$, where $L$ is the number of amino-acid sites, a given concatenated sequence can feature any amino acid (denoted by $\alpha$ with $\alpha\in\{1,..,20\}$), or a gap (denoted by $\alpha=21$), yielding $21$ possible states $\alpha$ for each site~$i$.

To describe the statistics of the alignment, we only employ the single-site frequencies of occurrence of each state $\alpha$ at each site $i$, denoted by $f^e_i(\alpha)$, and the  two-site frequencies of occurrence of each ordered pair of states $(\alpha,\beta)$ at each ordered pair of sites $(i,j)$, denoted by $f^e_{ij}(\alpha,\beta)$~\cite{Weigt09}. The raw empirical frequencies, obtained by counting the sequences where given residues occur at given sites and dividing by the number $M$ of sequences in the CA, are subject to sampling bias, due to phylogeny and to the choice of species that are sequenced~\cite{Morcos11,Marks11}. Hence, to define  $f^e_i$ and $f^e_{ij}$, we use a standard correction that re-weights ``neighboring'' concatenated sequences with mean Hamming distance per site $<0.3$. The value of this similarity threshold is arbitrary, but our results depend very weakly on this choice, even when taking the threshold down to zero. The weight associated to a given concatenated sequence $a$ is $1/m_a$, where $m_a$ is the number of neighbors of $a$ within the threshold~\cite{Marks11,Procaccini11,Morcos11}. This allows one to define an effective sequence number $M_\textrm{eff}$ via
\begin{equation}
M_\textrm{eff}=\sum_{a=1}^M \frac{1}{m_a}\,.
\label{Meff}
\end{equation}

To avoid issues such as amino acids that never appear at some sites, which would present mathematical difficulties, e.g. a non-invertible correlation matrix and diverging couplings~\cite{Morcos11}, we introduce pseudocounts via a parameter $\Lambda$~\cite{Weigt09,Procaccini11,Marks11,Morcos11}. The one-site frequencies $f_i$ become
\begin{equation}
f_i(\alpha)=\frac{\Lambda}{q}+(1-\Lambda)f^e_i(\alpha)\,,
\label{fi}
\end{equation}
where $q=21$ is the number of states (i.e. of amino acids, including gaps) per site. Similarly, the two-site frequencies $f_{ij}$ become
\begin{align}
f_{ij}(\alpha,\beta)&=\frac{\Lambda}{q^2}+(1-\Lambda)f^e_{ij}(\alpha,\beta)\textrm{ if }i\neq j\,, \label{fij}\\
f_{ii}(\alpha,\beta)&=\frac{\Lambda}{q} \delta_{\alpha\beta}+(1-\Lambda)f^e_{ii}(\alpha,\beta)= f_i(\alpha)\delta_{\alpha\beta}\,, \label{fii}
\end{align}
 where $\delta_{\alpha\beta}=1$ if $\alpha=\beta$ and 0 otherwise. These pseudocount corrections are uniform (i.e. they have the same weight $1/q$ on all amino-acid states), and their importance relative to the raw empirical frequencies can be tuned through the parameter $\Lambda$. In practice, we take $\Lambda=0.5$, which has been shown to be a satisfactory choice~\cite{Morcos11,Marks11}. Note that the correspondence of $\Lambda$ with the parameter $\lambda$ in Refs.~\cite{Procaccini11,Marks11,Morcos11} is obtained by setting $\Lambda=\lambda/(\lambda+M_\mathrm{eff})$.

From these quantities, we define the two-point correlations
\begin{equation}
C_{ij}(\alpha,\beta)=f_{ij}(\alpha,\beta)-f_i(\alpha)f_j(\beta)\,.
\label{Cij}
\end{equation}

\section{Maximum entropy model}

\subsection{Formulation}
The maximum entropy principle~\cite{Jaynes57} yields the following form for the least-structured global ($L$-point) probability distribution $P$ of sequences consistent with the empirical one- and two-point statistics of the CA:
\begin{equation}
P(\alpha_1,...,\alpha_L)=\frac{1}{Z}\exp\left\{-\left[\sum_{i=1}^{L}h_{i}(\alpha_i)+\sum_{i<j}e_{ij}(\alpha_i,\alpha_j)\right]\right\}\,,
\label{maxent}
\end{equation}
where $Z$ is a normalization constant. Each one-body term $h_i$ is known as the field at site $i$, and each two-body interaction term $e_{ij}$ is known as the (direct) coupling between sites $i$ and $j$. The fields $h_i$ and the couplings $e_{ij}$ are determined by imposing that the probability distribution $P$ be consistent with the empirical one- and two-point frequencies $f_i$ and $f_{ij}$:
\begin{align}
\sum_{\alpha_k,k\neq i} P(\alpha_1,...,\alpha_L)&=f_i(\alpha_i)\,, \label{consist1}\\
\sum_{\alpha_k,k\notin \{ i,j\}} P(\alpha_1,...,\alpha_L)&=f_{ij}(\alpha_i,\alpha_j)\,. \label{consist2}
\end{align}

Such pairwise interaction maximum entropy models have proved very successful in various fields (see e.g. Refs.~\cite{Schneidman06,Lezon06,Mora10,Bialek12,Wood12,Ferguson13,Mann14,Dwyer13,Figliuzzi16}), including the prediction of protein structures and inter-protein contacts from multiple sequence alignments (see e.g. Refs.~\cite{Weigt09,Morcos11,Marks11}). In particular, high couplings $e_{ij}$ are better predictors of real contacts in proteins than high correlations $C_{ij}$, because the $e_{ij}$ represent minimal direct couplings between amino acids, while high $C_{ij}$ can arise from indirect effects~\cite{Weigt09,Morcos11,Marks11}.

\subsection{Inference of the parameters} 
Eqs.~\ref{consist1} and~\ref{consist2} alone do not uniquely define all the fields $h_i(\alpha)$ and couplings $e_{ij}(\alpha,\beta)$ with $1\leq i<j\leq L$ involved in Eq.~\ref{maxent}, which amount to $Lq+L(L-1) q^2/2$ parameters, where $q=21$ is the number of amino-acid states $\alpha$. Indeed, while the number of equations in Eqs.~\ref{consist1} and~\ref{consist2} is the same as that of the empirical frequencies, the latter are not all independent. The two-site frequencies are symmetric ($f_{ij}(\alpha,\beta)=f_{ji}(\beta,\alpha)$) and consistent with the one-site frequencies ($f_{ii}(\alpha,\beta)=f_i(\alpha)\delta_{\alpha\beta}$;  $\sum_{\beta} f_{ij}(\alpha,\beta)=f_i(\alpha)$; and $\sum_{\alpha} f_{ij}(\alpha,\beta)=f_j(\beta)$), which sum to one ($\sum_\alpha f_i(\alpha)=1$). All these constraints reduce the number of independent variables among the one- and two-site frequencies, and thus of independent equations, to $L(q-1)+L(L-1) (q-1)^2/2$~\cite{Weigt09,Morcos11}. This yields a degree of freedom in the determination of the fields and couplings from Eqs.~\ref{consist1} and~\ref{consist2}. Given the number of independent equations, one possible gauge choice is to set to zero the fields and couplings for one given state, e.g. state $q$ (the gap)~\cite{Morcos11,Marks11}: $h_i(q)=0$ and, for all $\alpha$, 
\begin{equation}
e_{ij}(\alpha,q)=e_{ij}(q,\alpha)=0\,.
\label{gauge1}
\end{equation}

Determining the remaining fields $h_i$ and the couplings $e_{ij}$ from Eqs.~\ref{consist1} and~\ref{consist2} is difficult, and various approximations have been developed to solve this problem. Following Refs~\cite{Marks11, Morcos11}, we use the mean-field or small-coupling approximation, which was introduced in Ref.~\cite{Plefka82} for the Ising spin-glass model. In this approximation, for $i\neq j$ and $\alpha,\beta<q$, the couplings are given by $e_{ij}(\alpha,\beta)=A^{-1}_{kl}$, where $A$ is a $(q-1)L\times (q-1)L$ correlation matrix: $A_{kl}=C_{ij}(\alpha,\beta)$, where $k=(q-1)(i-1)+\alpha$ and $l=(q-1)(j-1)+\beta$~\cite{Ekeberg13}. 
This can be summarized as
\begin{equation}
e_{ij}(\alpha,\beta)=C^{-1}_{ij}(\alpha,\beta)\,.
\label{eij2}
\end{equation}
Together, Eqs.~\ref{gauge1} and~\ref{eij2} yield all the couplings. Note that the couplings are symmetric ($e_{ij}(\alpha,\beta)=e_{ji}(\beta,\alpha)$) since the correlations are. 

This simple mean-field approximation has been used with success for protein structure prediction~\cite{Morcos11,Marks11}. (More sophisticated approximations typically improve performance by less than ten percent~\cite{Ekeberg13,Baldassi14}.) Moreover, this approximation is computationally fast, since it only requires the inversion of a $(20 L)\times (20 L)$ correlation matrix. Computational rapidity is a considerable asset for our purpose, given that the IPA performs better with smaller increment step size $N_\mathrm{increment}$ (see Fig.~\ref{Fig3}), i.e. with more iterations, and that the couplings $e_{ij}$ are computed at each iteration. This approximation also enabled us to use the full-length sequences of domains to infer couplings, without needing to restrict to a subset of amino-acid sites as in some other works using more sophisticated approximations~\cite{Weigt09,Procaccini11}. We find that using full-length sequences increases the resulting TP fraction.

\subsection{Gauge choice}
Qualitatively, the gauge degree of freedom means that contributions to the effective energy of the system 
\begin{equation} 
H=\sum_{i=1}^{L}h_{i}(\alpha_i)+\sum_{i<j}e_{ij}(\alpha_i,\alpha_j)
\end{equation}
can be shifted between the fields and the couplings~\cite{Weigt09}. Since our focus is on interactions, we do not want the couplings to include contributions that can be accounted for by the (one-body) fields~\cite{Ekeberg14}. The zero-sum (or Ising) gauge, where the couplings satisfy
\begin{equation}
\sum_{\alpha}e_{ij}(\alpha,\beta)=\sum_{\beta}e_{ij}(\alpha,\beta)=0\,,
\label{gauge2}
\end{equation}
minimizes the Frobenius norms of the couplings
\begin{equation}
\left\Vert e_{ij}\right\Vert=\sqrt{\sum_{\alpha,\beta=1}^q \left[e_{ij}(\alpha,\beta)\right]^2}\,.
\label{Frob}
\end{equation}
Hence, the zero-sum gauge attributes the smallest possible fraction of the energy in Eq.~\ref{maxent} to the couplings, and the largest possible fraction to the fields~\cite{Weigt09,Ekeberg13}. Furthermore, when employing this gauge, the Frobenius norm has proved to be a successful predictor of contacts in proteins~\cite{Ekeberg13,Baldassi14}. In particular, within the mean-field approximation Eq.~\ref{eij2}, the use of the Frobenius norm (with an average-product correction) improves over the results obtained using direct information~\cite{Ekeberg13}. 

Thus, after calculating the couplings as described above, we change the gauge from the one defined in Eq.~\ref{gauge1} to the one defined in Eq.~\ref{gauge2}, by replacing each coupling $e_{ij}(\alpha,\beta)$ by
\begin{equation}
e_{ij}(\alpha,\beta)-\langle e_{ij}(\gamma,\beta)\rangle_\gamma-\langle e_{ij}(\alpha,\delta)\rangle_\delta+\langle e_{ij}(\gamma,\delta)\rangle_{\gamma,\delta}\,,
\label{gc}
\end{equation}
where $\langle .\rangle_\gamma$ denotes an average over $\gamma\in\{1,...,q\}$~\cite{Ekeberg13}.

Note that in Fig.~\ref{Fig4}, we use the Frobenius norm without the average-product correction~\cite{Ekeberg13}. With this correction, implemented by averaging within single proteins~\cite{Ovchinnikov14}, we obtained similar results (see Fig.~\ref{FigS_APC}). Overall, with the correction, final performance is slightly worse, but training is visible slightly earlier in the IPA.

\section{Iterative pairing algorithm (IPA)}

The main steps of the IPA are shown in Fig.~\ref{Fig1}C. Here, we describe each of these steps in detail, after explaining how the CA is constructed for the very first iteration. 

\subsection*{Initialization of the CA}
\subsubsection*{Starting from a training set of HK-RR pairs} The CA for the first iteration of the IPA is built from the pairs in the training set, which are considered as known interaction partners. In subsequent iterations, the training set pairs are \emph{always kept} in the CA, and additional pairs with the highest confidence scores (see below) are added to the CA.

\subsubsection*{Starting from random pairings} In the absence of a training set, each HK of the dataset is randomly paired with an RR from its species. All $M$ pairs, where $M$ represents the total number of HKs, or, equivalently, RRs, in the dataset, are included in the CA for the first iteration of the IPA. Hence, this initial CA contains a mixture of correct and incorrect pairs, with one correct pair per species on average. At the second iteration, the CA is built using only the $N_\mathrm{increment}$ HK-RR pairs with the highest confidence scores obtained from this first iteration. 

There are other ways to initialize the CA in the absence of a training set. We varied the number of pairs included at the second iteration ($N_\mathrm{increment}$ in the above scheme), and we also tried constructing the first CA from all possible HK-RR pairs from the species with few pairs (as for these species, exhaustive pairing yields a larger proportion of true pairs). These variants did not significantly increase the final TP fraction. Moreover, the random initialization of the CA can be exploited to increase the TP fraction (Figs.~\ref{FigS_NewPPI} and~\ref{FigS_Reboot}), which would be impossible for exhaustive initializations. 

Now that we have described the initial construction of the CA, we describe each step of an iteration of the IPA (Fig.~\ref{Fig1}C).

\subsection*{Step 1: Correlations}
At each iteration, the empirical one- and two-body frequencies are computed for the CA, using the re-weighting of neighbor sequences and the pseudocount correction described above (see Eqs.~\ref{Meff}-\ref{fii}). The empirical correlations $C_{ij}$ are then deduced using Eq.~\ref{Cij}.

\subsection*{Step 2: Direct couplings}
The direct couplings in the pairwise maximum entropy model of the CA are inferred from the empirical correlations using Eqs.~\ref{gauge1} and~\ref{eij2}. The gauge is then changed to the zero-sum gauge (Eq.~\ref{gauge2}) using Eq.~\ref{gc}.

\subsection*{Step 3: Interaction energies for all possible HK-RR pairs}
The interaction energy $E$ of each possible HK-RR pair within each species of the dataset is calculated by summing the appropriate direct couplings:
\begin{equation}
E\left(\alpha_1,...,\alpha_{L_\mathrm{HK}},\alpha_{L_\mathrm{HK}+1},...,\alpha_L\right)=\sum_{i=1}^{L_\mathrm{HK}}\sum_{j=L_\mathrm{HK}+1}^{L}e_{ij}(\alpha_i,\alpha_j)\,,
\end{equation}
where $L_\mathrm{HK}$ denotes the length (i.e. the number of amino-acid sites) of the HK sequence and $L$ that of concatenated HK-RR sequence. Note that this HK-RR interaction energy only involves the inter-molecular couplings ($i\leq L_\mathrm{HK}$ and $j> L_\mathrm{HK}$; the case $i> L_\mathrm{HK}$ and $j\leq L_\mathrm{HK}$ does not need to be considered as the couplings are symmetric).

\subsection*{Step 4: HK-RR pair assignments and ranking by energy gap}

\subsubsection*{HK-RR pair assignments} In each separate species, the pair with the lowest interaction energy is selected first, and the HK and RR from this pair are removed from further consideration, since we assume one-to-one HK-RR matches (see Fig.~\ref{Fig1}D). Then, the pair with the next lowest energy is chosen, and the process is repeated until all HKs and RRs are paired. 

\subsubsection*{Scoring by gap}
\label{SI_Meth_gap}
Each assigned HK-RR pair is scored at the time of assignment by $\Delta E/(n+1)$, where $\Delta E$ is the energy gap between the match with the lowest energy and the next best one (see Fig.~\ref{Fig1}E), and $n$ is the number of lower-energy matches discarded in assignments made previously (within that species and at that iteration). Qualitatively, the larger the energy gap, and the smaller the number $n$ of rejected better candidates, the more reliable we expect the assignment to be.

More precisely, $\Delta E_\textrm{RR}=E_\textrm{RR,2}-E_\textrm{RR,1}>0$ is computed for the RR involved as minus the difference of the interaction energy $E_\textrm{RR,1}$ of this RR with its assigned partner (i.e. the ``best'' HK, which yields the lowest interaction energy with this RR,  \emph{among the HKs that are still unpaired}) and that $E_\textrm{RR,2}$ with the second-best HK \emph{among the HKs that are still unpaired}. Meanwhile, $n_\textrm{RR}$ is the number of HKs of that species that had lower interaction energies with this RR than the assigned partner, but that have been eliminated previously in that iteration's pairing process, because they were paired to other RRs with a lower interaction energy. A schematic example is shown on Fig.~\ref{FigS_gap}A. Similarly, the value of $\Delta E_\textrm{HK}$ and of $n_\textrm{HK}$ are calculated for the HK involved in the assigned pair. Finally, the lowest score among the two obtained is kept:
\begin{equation}
\frac{\Delta E}{n+1}=\min\left(\frac{\Delta E_\textrm{RR}}{n_\textrm{RR}+1},\,\,\frac{\Delta E_\textrm{HK}}{n_\textrm{HK}+1}\right)\,.
\end{equation} 

 We have chosen to divide the energy gap $\Delta E$ by $n+1$ in order to penalize the HK-RR pairs made after better candidates were discarded, even if their current gap among remaining candidates appears large, as illustrated by the second assignment in Fig.~\ref{FigS_gap}A. However, one could consider other definitions of confidence scores, such as $\Delta E/(n+1)^\alpha$, where $\alpha$ is a parameter. In Fig.~\ref{FigS_gap}B, we show that our confidence score significantly improves TP fraction over the raw energy gap $\Delta E$, and that $\alpha=1$ yields an optimal TP fraction.
 
This definition of the confidence score leaves an ambiguity for the last assigned pair of each species, since there is no remaining second-best match to define the energy gap. We have chosen to assign to this pair a confidence score equal to the lowest other one within the species, given that this pair, made by default, should not be deemed more reliable than any other pair in the species. 

Another ambiguity exists when several pairs have exactly the same interaction energy. This mostly occurs when the model is built from one single HK-RR concatenated sequence (this case is not singular thanks to the pseudocount correction, and the model then yields a lower energy contribution for each residue pair identical to the initial concatenated sequence, and a higher energy contribution for each residue pair comprising one same and one different residue compared to the initial concatenated sequence). It also occurs in the extremely rare case where two identical HK (or RR) sequences are found in the same genome. In this case, we chose to randomly make one pair assignment between the equivalent matches, and to leave the other equal energy HKs and/or RRs to be paired later. We checked that the impact of this choice on final results is very small.

\subsubsection*{Ranking of pairs} 
Once all the HK-RR pairs are assigned and scored, we rank them in order of decreasing confidence score. 
 
 \subsection*{Step 5: Incrementation of the CA}
The ranking of the HK-RR pairs is used to pick those pairs that are included in the CA at the next iteration. Pairs with a high confidence score are more likely to be correct because there was less ambiguity in the assignment. The number of pairs in the CA is increased by $N_\mathrm{increment}$ at each iteration, and the IPA is run until all the HKs and RRs in the dataset have been paired and added to the CA. In the last iteration, all pairs assigned at the second to last iteration are included in the CA. 

\subsubsection*{Starting from a training set of HK-RR pairs} The $N_\mathrm{start}$ training pairs remain in the CA throughout and the HKs and RRs involved in these pairs are not paired or scored by the IPA. The HKs and RRs from all the other pairs in the CA are re-paired and re-scored at each iteration, and only re-enter the CA if their confidence score is sufficiently high. In other words, at the first iteration, the CA only contains the $N_\mathrm{start}$ training pairs. Then, for any iteration number $n>1$, it contains these exact same $N_\mathrm{start}$ training pairs, plus the $(n-1)N_\mathrm{increment}$ assigned HK-RR pairs that had the highest confidence scores at iteration number $n-1$. 

\subsubsection*{Starting from random pairings} In the absence of a training set, all $M$ HKs and RRs in the dataset are paired and scored at each iteration, and all the pairs of the CA are fully re-picked at each iteration based on the confidence score. The first iteration is special, since the CA is made of $M$ random within-species HK-RR pairs (see above, ``Initialization of the CA''). Then, for any iteration number $n>1$, the CA contains the $(n-1) N_\mathrm{increment}$ assigned HK-RR pairs that had the highest confidence scores at iteration number $n-1$. 

Once the new CA is constructed, the iteration is completed, and the next one can start with Step 1, the computation of the empirical correlations in this CA.

\subsection*{Run time} 

The run time of the IPA strongly depends on $N_\mathrm{increment}$ and on dataset size (length of concatenated sequences, number of such sequences in the dataset). For our standard HK-RR dataset, all single-processor run times for a Matlab-coded version of the IPA were shorter than one day down to $N_\mathrm{increment}=6$.

\clearpage
\onecolumngrid

\centerline{\textbf{\textsf{SUPPORTING FIGURES}}}
\vspace{1cm}

\begin{figure}[hb]
\centering
\includegraphics[width=.4\linewidth]{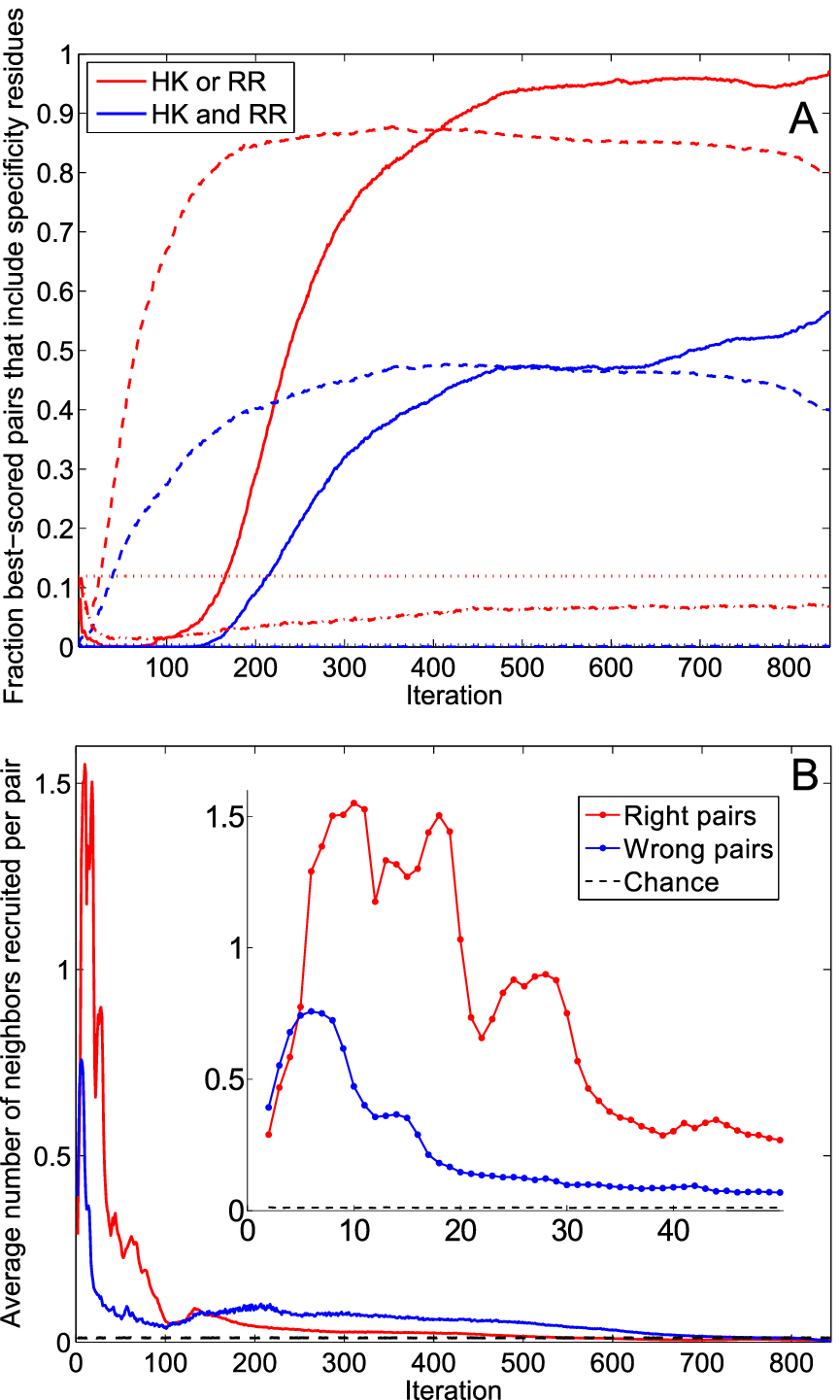}
\caption{Evolution of the coupling matrix and of the concatenated alignment (CA) during the IPA. (A) Training of the coupling matrix. As in Fig.~\ref{Fig4}A, pairs comprised of an HK residue site and an RR residue site are scored by the Frobenius norm (i.e. the square root of the summed squares) of the couplings involving all possible residue types at these two sites. The 10 best-scored pairs are compared to the main specificity residues determined experimentally in Refs.~\cite{Skerker08,Capra10,Podgornaia13b,Podgornaia15} (5 HK residues, T267, A268, A271, Y272, and T275 in the sequence of \textit{T. maritima} HK853, and 5 RR residues, V13, L14, I17, N20, and F21 in the sequence of \textit{T. maritima} RR468 ~\cite{Podgornaia13b}). Solid curves: Fraction of the 10 best-scored residue pairs that include HK and/or RR specificity residues versus the iteration number in the IPA. Dashed curves: Ideal case, where at each iteration $N_\mathrm{increment}$ randomly-selected correct HK-RR pairs are added to the CA. Dash-dotted curves: Case where random HK-RR pairs are added to the CA. Dotted lines: Overall fraction of residue pairs that include specificity residues. (B) Neighbor recruitment. Average number of neighbors an HK-RR pair of the CA has among the new HK-RR pairs of the next CA versus iteration number. Two pairs are considered neighbors if the mean Hamming distance per site between the two HKs and between the two RRs are both $<0.3$. Dashed line: Null model -- at each iteration, $N_{\mathrm{increment}}$ new correct HK-RR pairs are chosen at random and added to the CA. Inset: Expanded view of the first 50 iterations. In both panels, the IPA is performed on the standard dataset with $N_{\mathrm{increment}}=6$. In panel A (resp. B), data is averaged over 500 (resp. 5193) replicates that differ in their initial random pairings.}
\label{FigS_Pred}
\end{figure}

\begin{figure}
\centering
\includegraphics[width=.4\linewidth]{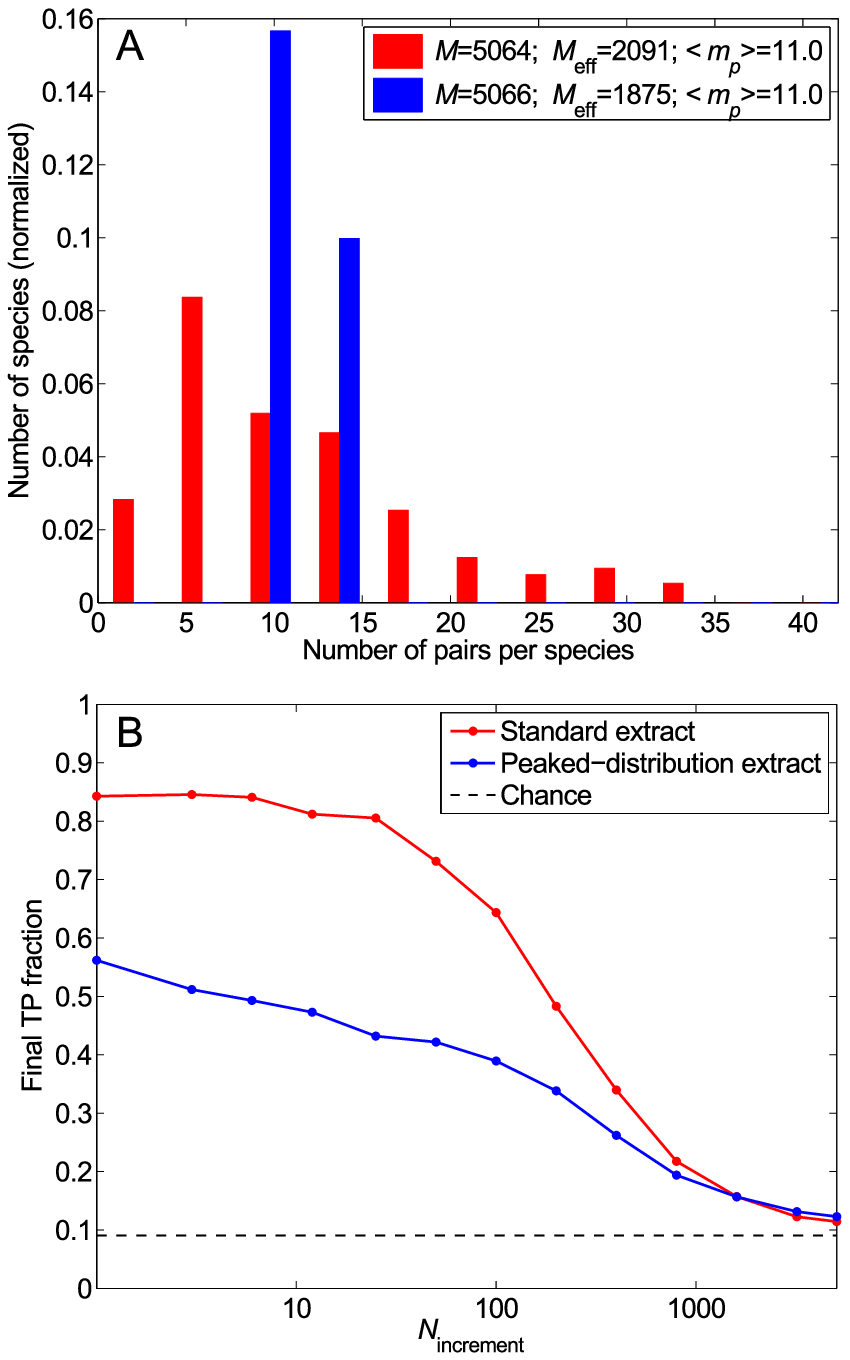}
\caption{Impact of the distribution of the number of HK-RR pairs per species. (A) Distribution of the number of pairs per species in two different datasets: the standard one (red) and one with the same total number of HK-RR pairs $M$ and the same mean number of pairs per species $\langle m_p\rangle$, but with a more strongly peaked distribution (blue). (B) Final TP fraction versus $N_{\mathrm{increment}}$ for the two datasets described in (A). All results are averaged over 50 replicates that differ in their initial random pairings. Dashed line: Average TP fraction obtained for random HK-RR pairings. }
\label{FigS_ImpFewPairs}
\end{figure}

\begin{figure}[b]
\centering
\includegraphics[width=.4\linewidth]{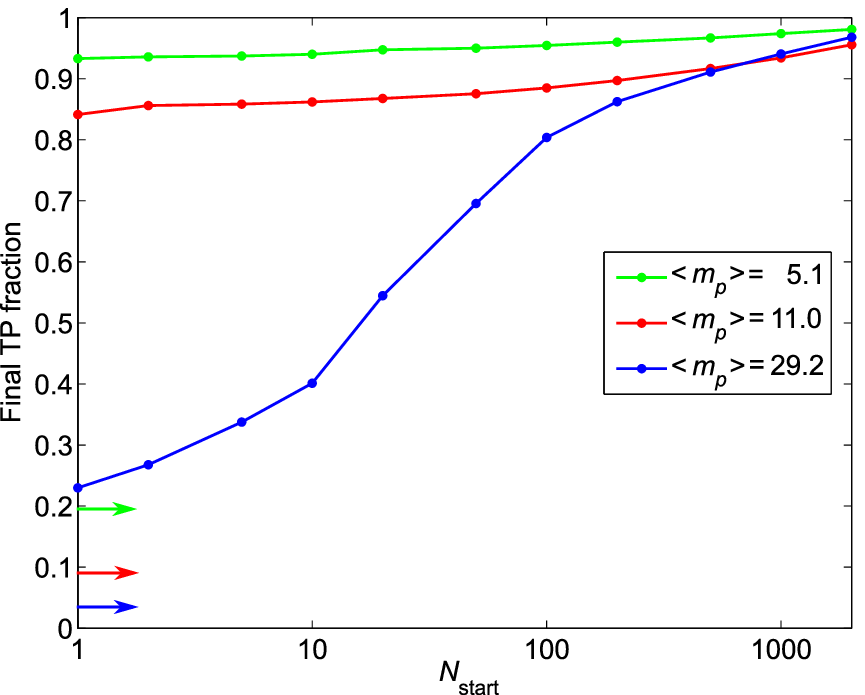}
\caption{Impact of the number of HK-RR pairs per species: starting from a training set. Final TP fraction versus $N_{\mathrm{start}}$ for the three datasets with different distributions of the number of pairs per species yielding different means $\langle m_p\rangle$ presented in Fig.~\ref{Fig5}. Colored arrows indicate the average TP fractions obtained for random HK-RR pairings in each dataset. The IPA is performed on the standard dataset with $N_{\mathrm{increment}}=6$. All results are averaged over 50 replicates that differ by the random choice of pairs in the training set.}
\label{FigS_TP_PPS}
\end{figure}

\begin{figure}
\centering
\includegraphics[width=.4\linewidth]{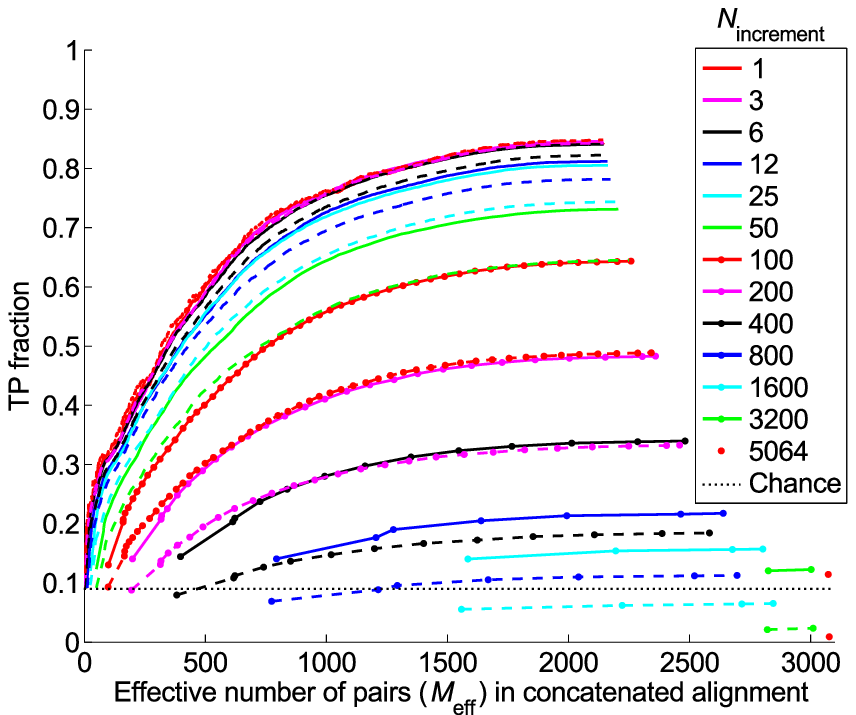}
\caption{Impact of the initial correct pairs. TP fraction versus effective number of HK-RR pairs ($M_{\mathrm{eff}}$) in the concatenated alignment during iterations of the IPA, for different values of $N_{\mathrm{increment}}$. Solid curves: Starting from random pairings (data also shown in Fig.~\ref{Fig3}). Dashed curves: Starting from random pairings with no initial correct pair (the color and symbol codes are the same as for the solid curves). The standard dataset is used. All results are averaged over 50 replicates that differ in their initial random pairings. Dotted line: Average TP fraction obtained for random HK-RR pairings.}
\label{FigS_IniImpact}
\end{figure}

\begin{figure}
\centering
\includegraphics[width=.4\linewidth]{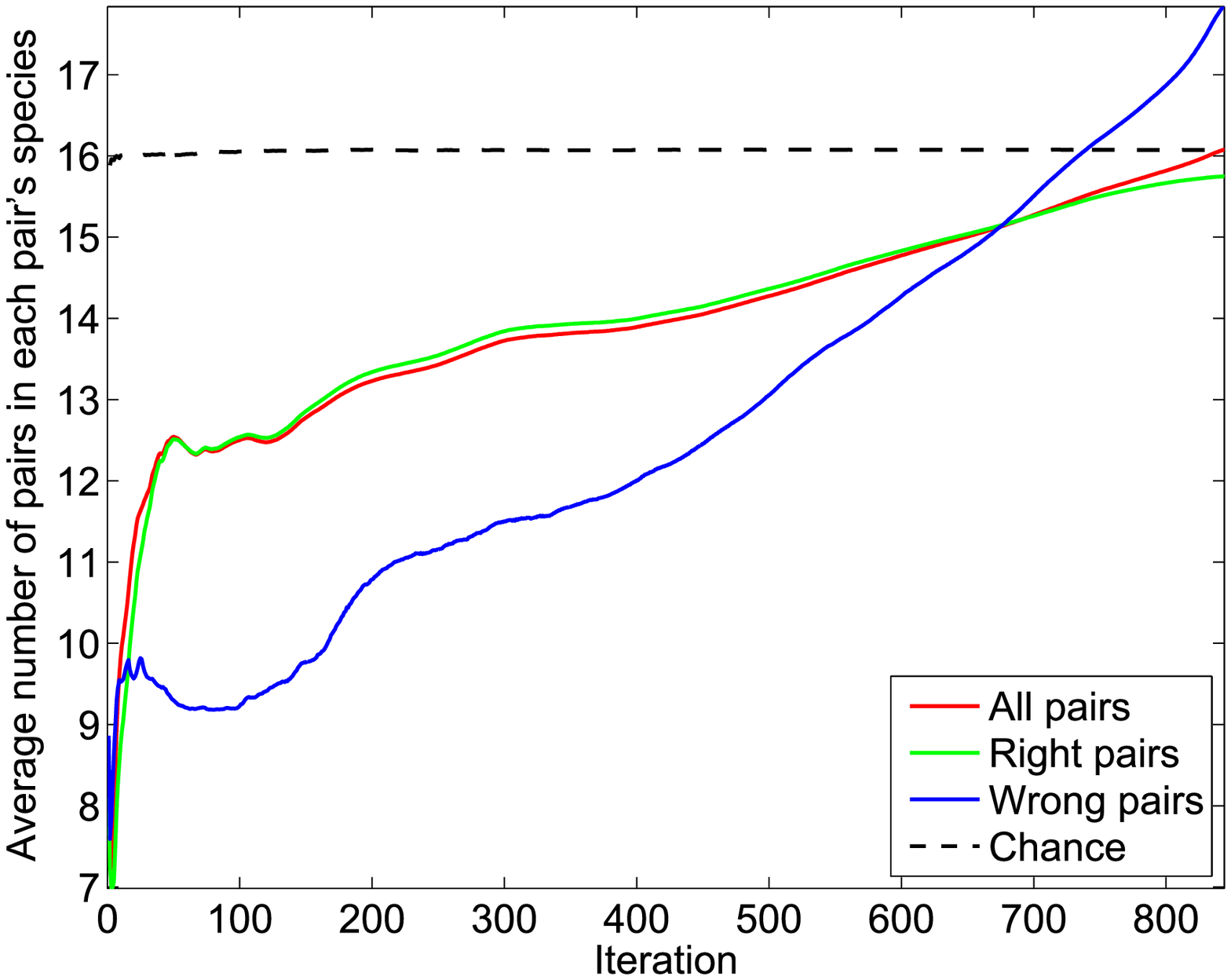}
\caption{Evolution of the concatenated alignment (CA) during the IPA. Average number of HK-RR pairs present in the species to which the pairs of the CA belong versus iteration number. The IPA is performed on the standard dataset, with $N_{\mathrm{increment}}=6$, and all data is averaged over 5193 replicates that differ in their initial random pairings. Dashed line: At each iteration, 6 new correct HK-RR pairs are chosen at random and added to the CA. This “chance” result just matches the average number of pairs in a pair's species: 16.1. Note that this number is different from the above-discussed average number of pairs per species $\langle m_p\rangle$, which is 11.0 in the standard dataset (because the average over the pairs is not the same as the average over the species).}
\label{FigS_Evol_PPS}
\end{figure}

\begin{figure}
\centering
\includegraphics[width=.4\linewidth]{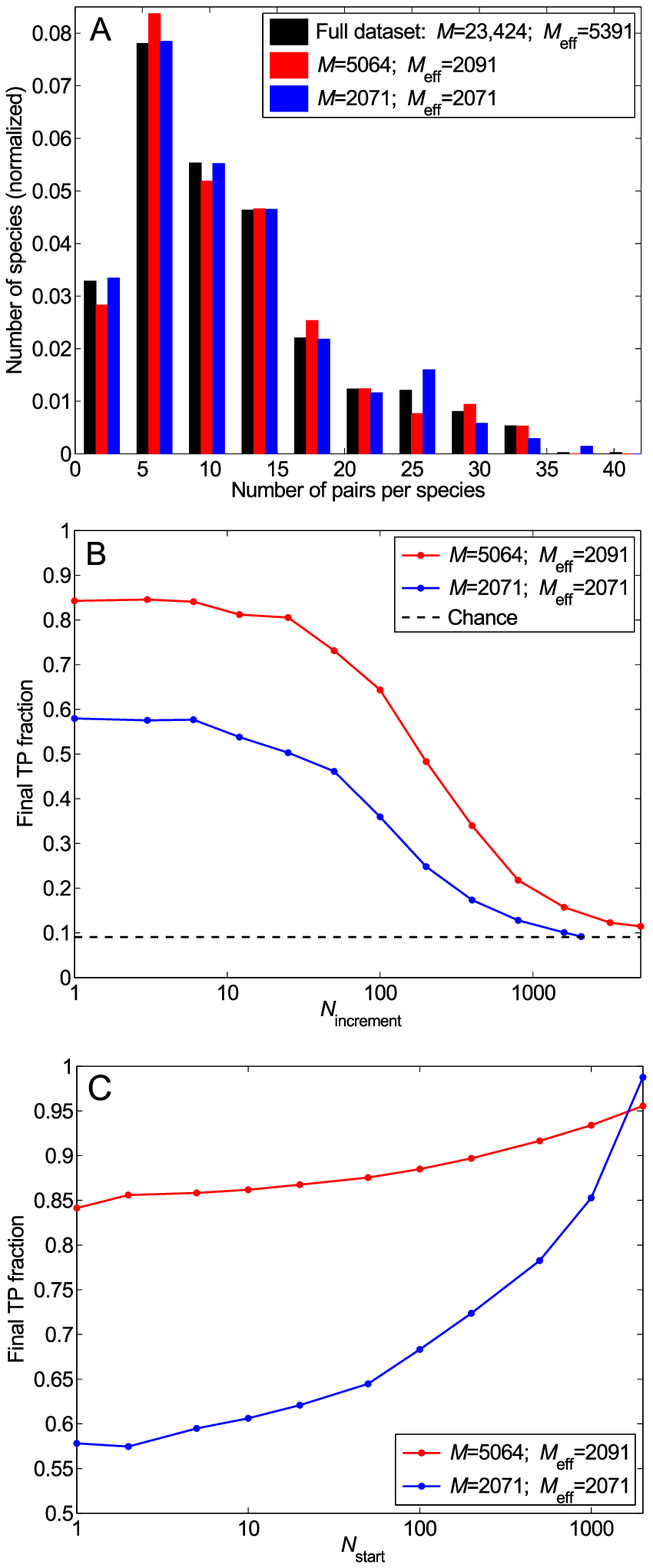}
\caption{Impact of sequence similarity in the dataset. (A) Distribution of the number of pairs per species in the complete dataset (black) and in two smaller selected datasets each with the same effective number of HK-RR pairs $M_\mathrm{eff}$: the standard one (red) and one where similar sequences have been suppressed such that no two pairs have a mean Hamming distance per site $<0.3$ (blue). (B) Final TP fraction versus $N_{\mathrm{increment}}$ for the two selected datasets described in (A), starting from random pairings. Dashed line: Average TP fraction obtained for random HK-RR pairings. (C) Starting from a training set. Final TP fraction versus $N_{\mathrm{start}}$ for the two selected datasets presented in (A), with $N_{\mathrm{increment}}=6$. In (B) and (C), all results are averaged over 50 replicates.}
\label{FigS_SeqSim}
\end{figure}

\begin{figure}
\centering
\includegraphics[width=.4\linewidth]{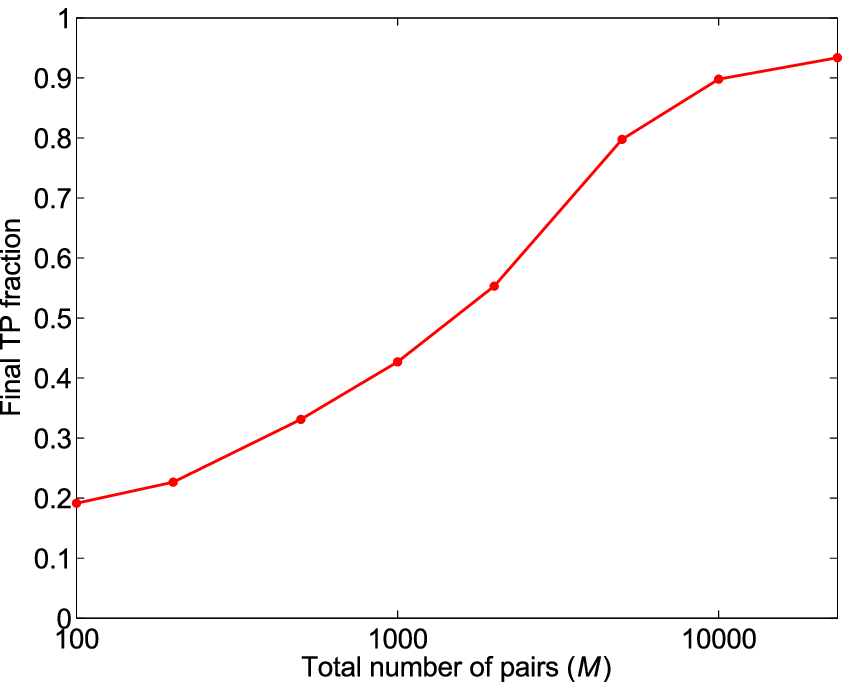}
\caption{Impact of the total number of HK-RR pairs in the dataset. Final TP fraction versus the total number $M$ of HK-RR pairs in the dataset, starting from random pairings. For each $M$, datasets are constructed by picking species randomly from the full dataset, preserving the average distribution of the number of HK-RR pairs per species. For each $M$ except the largest, results are averaged over multiple different such alignments (from 50 up to 500 for small $M$). For the largest $M$ (full dataset), averaging is done on 50 different initial random pairings. All results correspond to the small-$N_{\mathrm{increment}}$ limit.}
\label{FigS_NumberSeqs}
\end{figure}

\begin{figure}
\centering
\includegraphics[width=.4\linewidth]{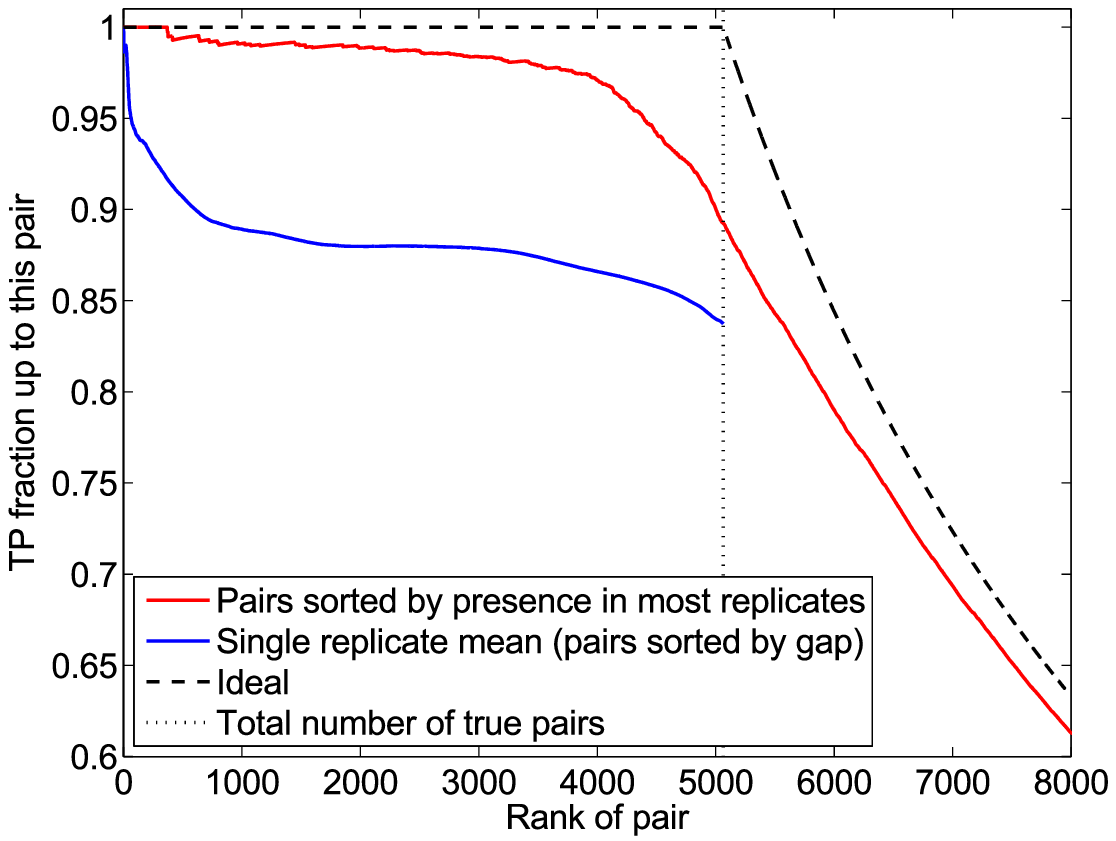}
\caption{Improved accuracy from multiple initial random pairings. Red curve: All possible HK-RR pairs (within each species) are ranked by the fraction $f_r$ of replicates of the IPA in which they are predicted. The TP fraction up to each pair is plotted versus the rank of this pair. The standard dataset is used, with $N_{\mathrm{increment}}=6$. 500 replicates that differ in their initial random pairings are considered. Blue curve: For each separate replicate, pairs are ranked by their confidence score, in decreasing order. The TP fraction up to each pair is computed, and the mean of these curves is shown. Dashed curve: Ideal classification, where the $M=5064$ first pairs (dotted line) are correct, while all the others are incorrect. When ranking pairs by decreasing $f_r$ (red curve), the TP fraction among the $M=5064$ best-ranked pairs is 0.89, a significant improvement over the average of TP fractions from individual replicates, 0.84 (blue curve).}
\label{FigS_RepFracRank}
\end{figure}

\begin{figure}
\centering
\includegraphics[width=.4\linewidth]{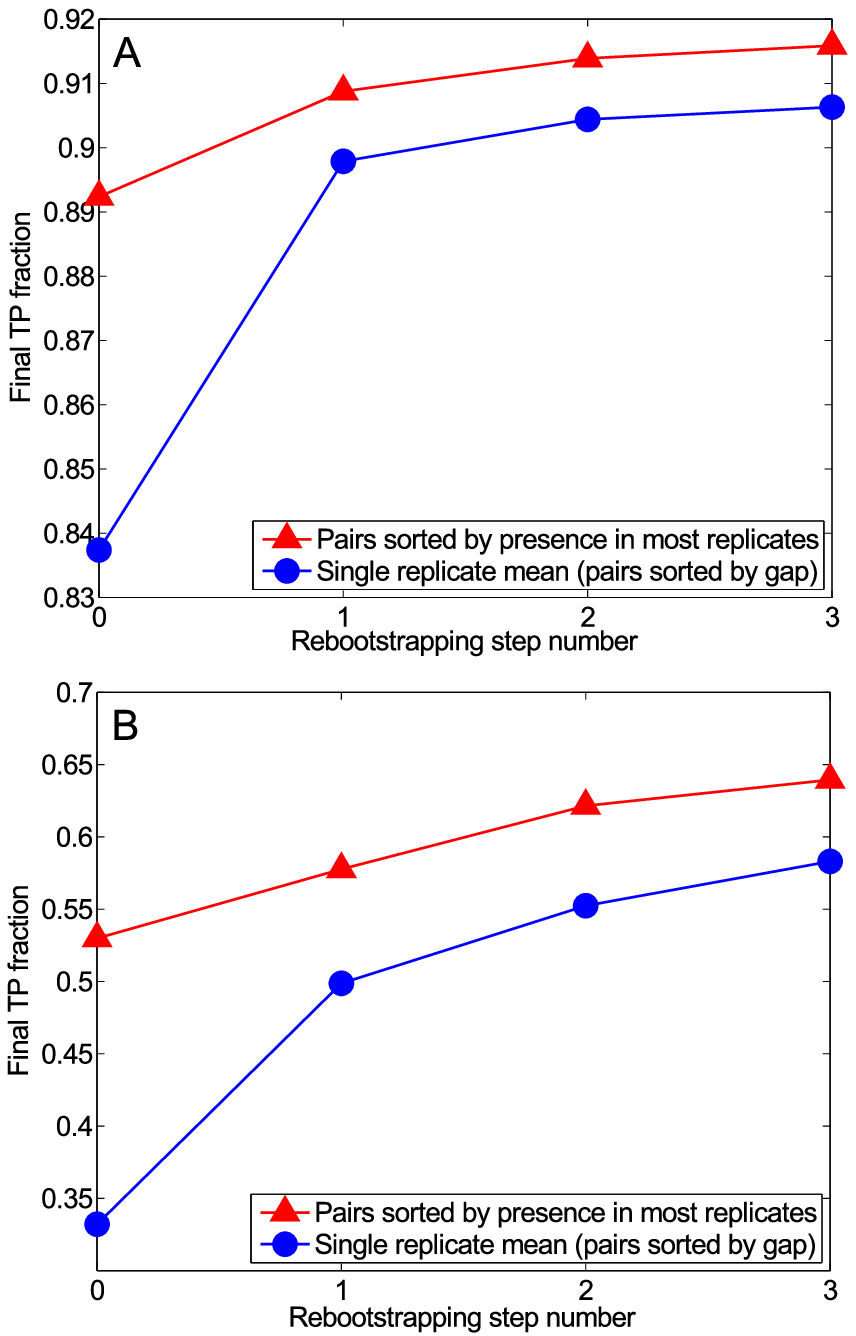}
\caption{Rebootstrapping: exploiting the high TP fraction of the HK-RR pairs predicted to be correct in most replicates of the IPA, which differ in their initial random pairings. (A) Rebootstrapping on the standard dataset ($M=5064$ HK-RR pairs). The final TP fraction is plotted versus rebootstrapping step number. Step 0 corresponds to the standard procedure described in the main text (IPA starting from random pairings, see Fig.~\ref{Fig6}). 500 replicates are computed. We then take as a training set 1000 HK-RR pairs chosen randomly among those predicted to be correct in more than 50\% of replicates. These pairs are chosen with probability equal to the fraction of replicates in which they are predicted to be true. The IPA is then performed again starting from such training sets. The process is then iterated. Here, 50 replicates were computed for steps 1, 2, and 3. The average final TP fraction is plotted (blue curve), as well as the TP fraction for the best $M=5064$ pairs ranked by the fraction of replicates in which they are predicted to be true (red curve, see Fig.~\ref{Fig6}). Here, $N_{\mathrm{increment}}=6$. (B) Rebootstrapping on a smaller dataset with $M=502$ HK-RR pairs from 40 species (mean number of pairs per species $\left<m_p \right>=12.6$). The process is the same as in (A), but here, at each rebootstrapping step, we take as a training set 200 HK-RR pairs chosen randomly among those predicted to be true in more than 25\% of replicates at the previous step, and $N_{\mathrm{increment}}=1$.}
\label{FigS_Reboot}
\end{figure}

\begin{figure}
\centering
\includegraphics[width=.4\linewidth]{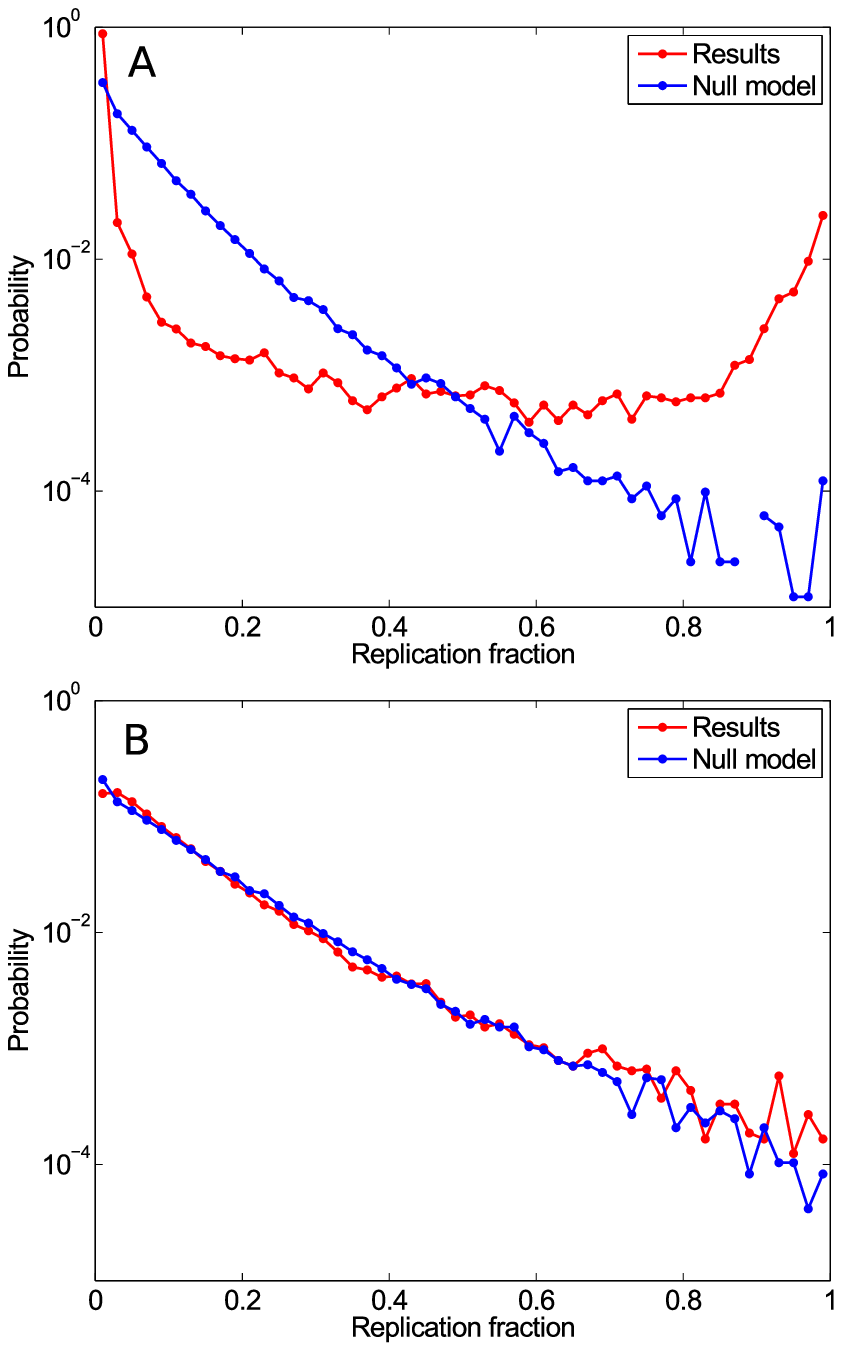}
\caption{Distribution of the fraction of replicates $f_r$ of the IPA in which each possible within-species HK-RR pair is predicted as a pair. (A) Red curve: Distribution of $f_r$ obtained by applying the IPA to the standard dataset (same data as in Fig.~\ref{FigS_RepFracRank}). Blue curve: Same dataset, but with each column of the alignment randomly scrambled. (B) HK-RR dataset with no correct pairs; a dataset of the same size as the standard one ($M=5062$ in practice) that does not include any true HK-RR pairs was constructed. Red curve: Distribution of $f_r$ obtained by applying the IPA to this dataset with no correct pairs. Blue curve: Same alignment, but with each column randomly scrambled. For each curve, 500 IPA replicates that differ in their initial random pairings were used, with $N_{\mathrm{increment}}=6$. All data is binned into 50 equally-spaced bins between $f_r=0$ and $f_r=1$. }
\label{FigS_NewPPI}
\end{figure}

\begin{figure}
\centering
\includegraphics[width=.4\linewidth]{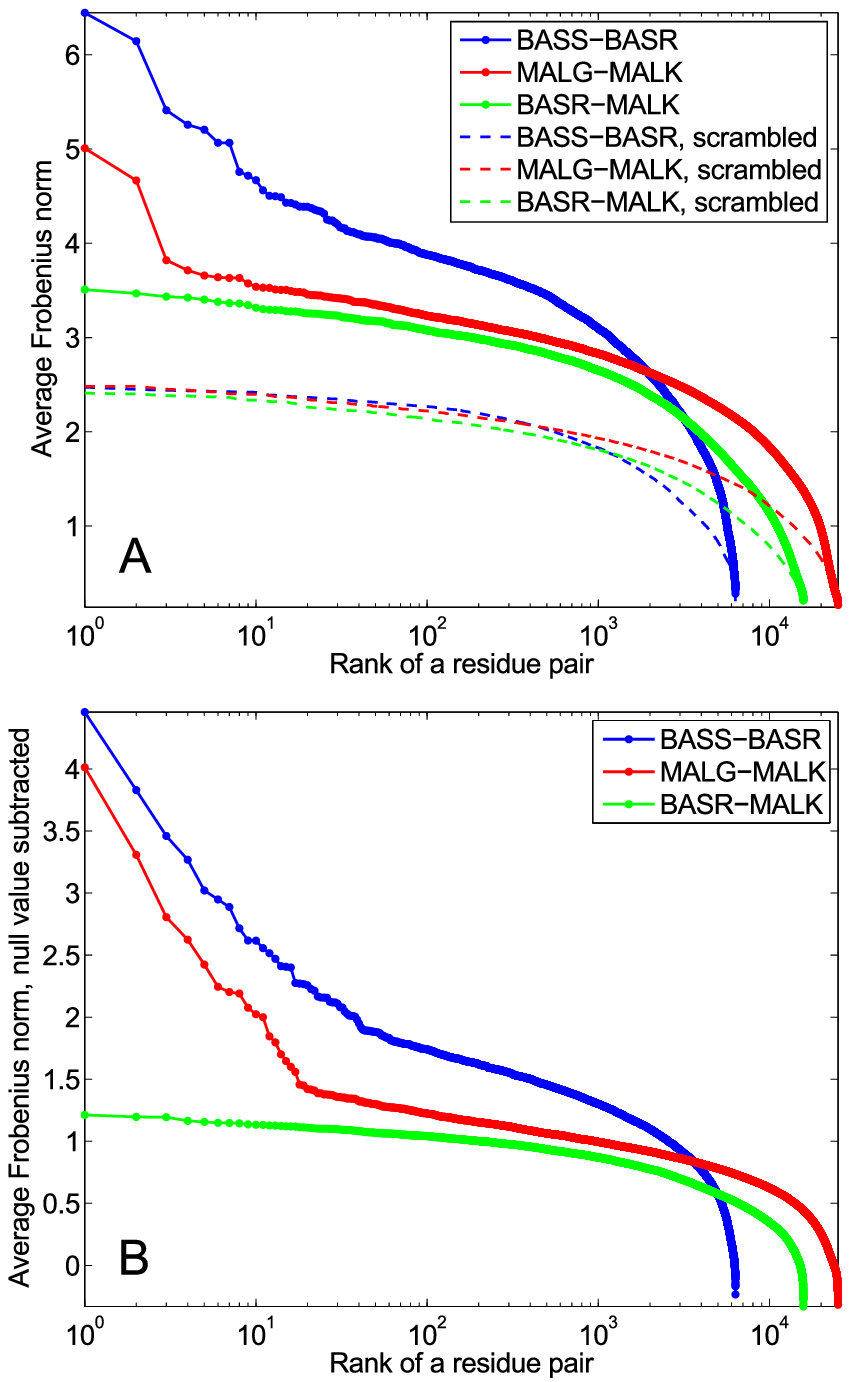}
\caption{ Residue-based signature of protein-protein interactions. The Frobenius norm of the amino-acid couplings was evaluated for each pair of residue sites at the final iteration of the IPA, for datasets comprising $\sim$5000 homologs of the interacting pairs BASS-BASR and MALG-MALK, and of the non-interacting pair BASR-MALK. For each of these protein family pairs, the Frobenius norms were also calculated at the final iteration of the IPA on scrambled-within-column datasets (null model). (A) Frobenius norms averaged over 500 IPA replicates that differ in their initial random pairings, and then ranked by decreasing value. (B) Same average Frobenius norms, normalized by subtracting the average null value for each residue pair. For each curve, the IPA was run with $N_{\mathrm{increment}}=50$. The pairs with highest Frobenius norms, corresponding to the top predicted contacts, are outliers for both interacting family pairs, but not for the non-interacting pair BASR-MALK.}
\label{FigS_NewPPI_Contacts}
\end{figure}

\begin{figure}
\centering
\includegraphics[width=.4\linewidth]{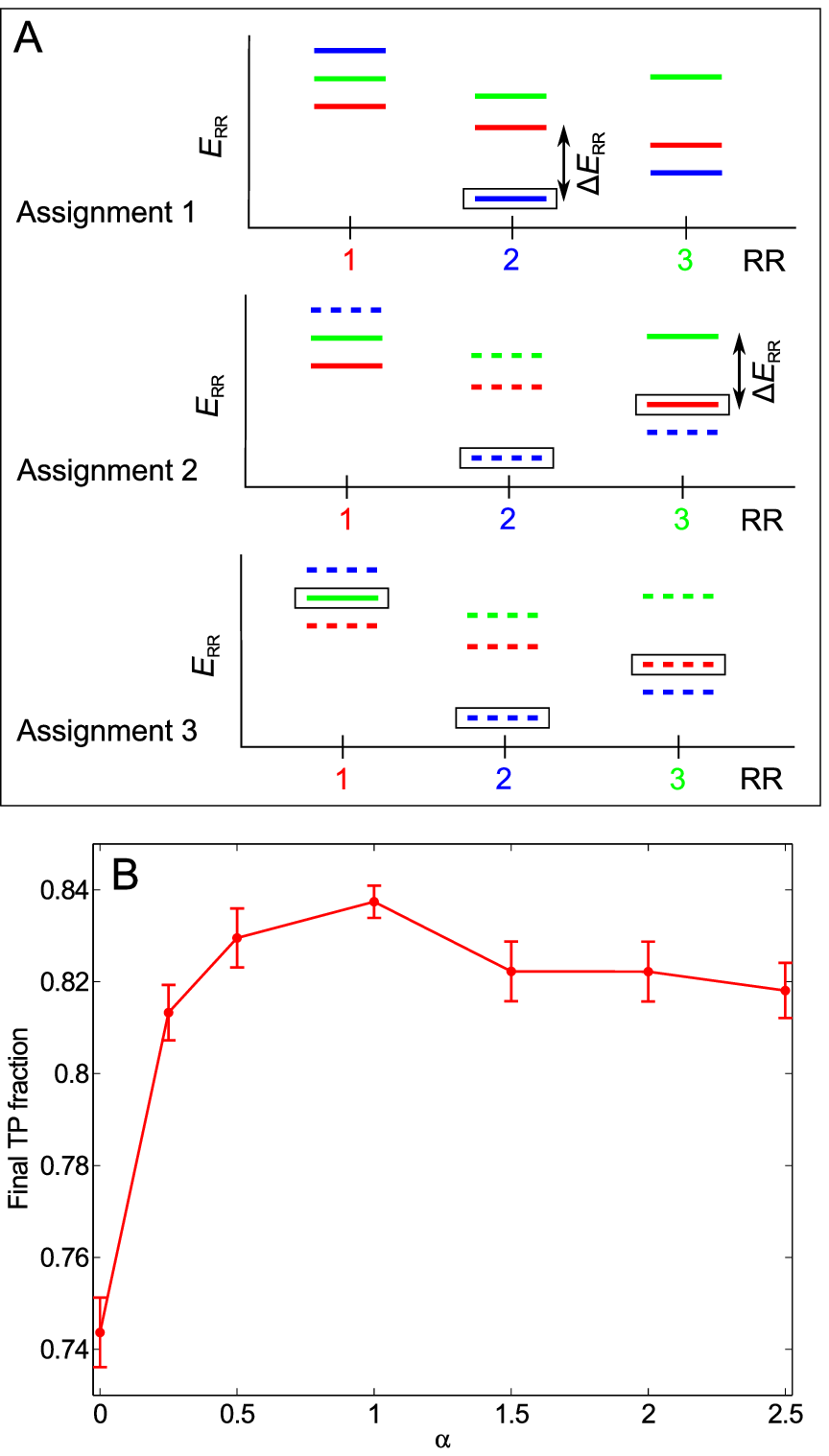}
\caption{Scoring by gap. (A) Determination of the confidence score of each assigned HK-RR pair in a given iteration of the IPA. In this schematic, we consider a species with three HKs and RRs. In the energy spectra showing the interaction energies for each RR with all three HKs, each color represents a given HK (red: HK 1, partner of RR 1; blue: 2; green: 3). Assignment 1: The pair with the lowest interaction energy (HK 2 - RR 2, boxed) is selected. The energy gap $\Delta E_\mathrm{RR}$ is shown. Here $n_\mathrm{RR}=0$ since no HK has been removed from consideration yet. Assignment 2:  The HK and RR previously paired are removed from further consideration (dashed energy levels). The next pair with the lowest energy (HK 1 - RR 3, boxed) is chosen among the remaining ones. Here $n_\mathrm{RR}=1$ since HK 2, which was paired previously, had a lower interaction energy with RR 3 than HK 1. Using the \textit{ad hoc} confidence score $\Delta E_\mathrm{RR}/(n_\mathrm{RR}+1)$, this (incorrect) pair is penalized with respect to the (correct) one made in the first assignment, even though their energy gaps are similar. Assignment 3: Only one possible pair remains. It is made, and its confidence score is taken to be equal to the lowest previously calculated confidence score for that species (the second one here). At each HK-RR pair assignment, symmetric confidence scores $\Delta E_\mathrm{HK}/(n_\mathrm{HK}+1)$ are also calculated from the energy spectra showing the interaction energies for each HK with all three RRs. The final confidence score of a pair, denoted by $\Delta E/(n+1)$, is the smallest of these two scores, i.e. $\min\{\Delta E_\mathrm{RR}/(n_\mathrm{RR}+1),\,\,\Delta E_\mathrm{HK}/(n_\mathrm{HK}+1)\}$. (B) More generally, in every iteration of the IPA, each predicted HK-RR pair can be scored by $\Delta E/(n+1)^\alpha$, where $\alpha$ is a parameter. Red curve: Average final TP fraction obtained versus $\alpha$; error bars: 95\% confidence intervals around the mean. The IPA was performed on the standard dataset, with $N_{\mathrm{increment}}=6$. Results are averaged over 200 replicates that differ in their initial random pairings for all $\alpha$ except $\alpha=1$, for which 500 replicates were computed. As we found the highest TP fraction for $\alpha = 1$, all the results elsewhere in the paper were obtained using $\alpha=1$.}
\label{FigS_gap}
\end{figure}

\begin{figure}
\centering
\includegraphics[width=.4\linewidth]{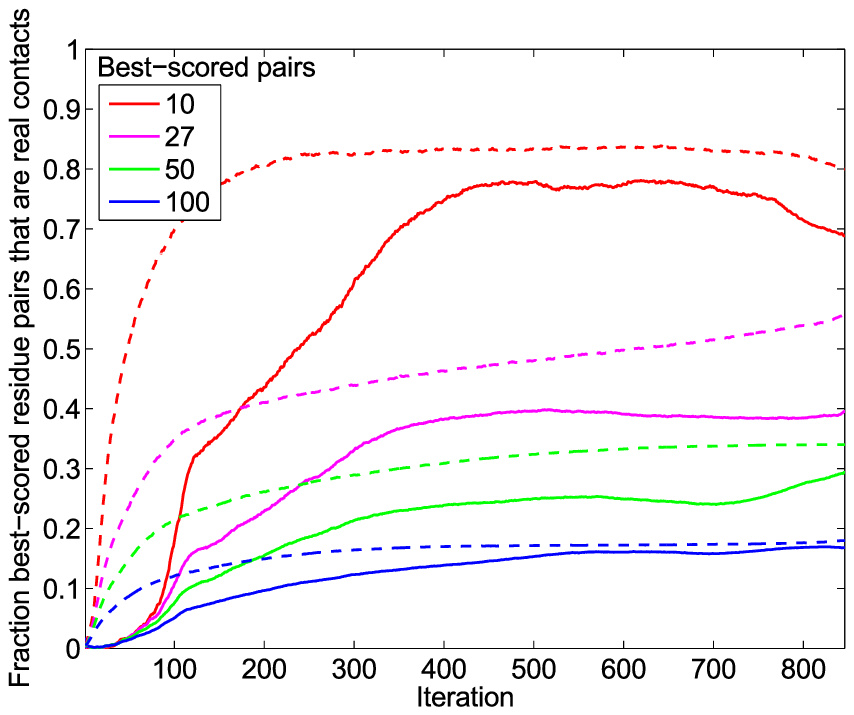}
\caption{Training of the couplings during the IPA: effect of the average product correction. Residue pairs comprised of an HK site and an RR site were scored by the average-product corrected Frobenius norm of the couplings involving all possible residue types at these two sites. The best-scored residue pairs were compared to the 27 HK-RR contacts found experimentally in Ref.~\cite{Casino09}. Solid curves: Fraction of residue pairs that are real contacts (among the $k$ best-scored pairs for four different values of $k$) versus the iteration number in the IPA. Dashed curves: Ideal case, where at each iteration $N_\mathrm{increment}$ randomly-selected correct HK-RR pairs are added to the CA. The overall fraction of residue pairs that are real HK-RR contacts, yielding the chance expectation, is only $3.8\times10^{-3}$. As in Fig.~\ref{Fig4}, the IPA was performed on the standard dataset with $N_{\mathrm{increment}}=6$, and all data is averaged over 500 replicates that differ in their initial random pairings.}
\label{FigS_APC}
\end{figure}


\begin{thebibliography}{10}

\bibitem{Rajagopala14}
Rajagopala SV et~al. (2014) {{T}he binary protein-protein interaction landscape
  of {E}scherichia coli}.
\newblock {\em Nat. Biotechnol.} 32(3):285--290.

\bibitem{Altschuh87}
Altschuh D, Lesk AM, Bloomer AC, Klug A (1987) {{C}orrelation of co-ordinated
  amino acid substitutions with function in viruses related to tobacco mosaic
  virus}.
\newblock {\em J. Mol. Biol.} 193(4):693--707.

\bibitem{Lockless99}
Lockless SW, Ranganathan R (1999) {{E}volutionarily conserved pathways of
  energetic connectivity in protein families}.
\newblock {\em Science} 286(5438):295--299.

\bibitem{Skerker08}
Skerker JM et~al. (2008) {{R}ewiring the specificity of two-component signal
  transduction systems}.
\newblock {\em Cell} 133(6):1043--1054.

\bibitem{Lapedes99}
Lapedes AS, Giraud BG, Liu L, Stormo GD (1999) Correlated mutations in models
  of protein sequences: phylogenetic and structural effects in {\em Statistics
  in molecular biology and genetics - IMS Lecture Notes - Monograph Series}.
\newblock Vol.{}~33, pp. 236--256.

\bibitem{Burger08}
Burger L, van Nimwegen E (2008) {{A}ccurate prediction of protein-protein
  interactions from sequence alignments using a {B}ayesian method}.
\newblock {\em Mol. Syst. Biol.} 4:165.

\bibitem{Weigt09}
Weigt M, White RA, Szurmant H, Hoch JA, Hwa T (2009) {{I}dentification of
  direct residue contacts in protein-protein interaction by message passing}.
\newblock {\em Proc. Natl. Acad. Sci. U.S.A.} 106(1):67--72.

\bibitem{Jaynes57}
Jaynes ET ({1957}) {Information Theory and Statistical Mechanics}.
\newblock {\em {Phys. Rev.}} {106}({4}):{620--630}.

\bibitem{Marks11}
Marks DS et~al. (2011) {{P}rotein 3{D} structure computed from evolutionary
  sequence variation}.
\newblock {\em PLoS ONE} 6(12):e28766.

\bibitem{Sulkowska12}
Su{\l}kowska JI, Morcos F, Weigt M, Hwa T, Onuchic JN (2012) {{G}enomics-aided
  structure prediction}.
\newblock {\em Proc. Natl. Acad. Sci. U.S.A.} 109(26):10340--10345.

\bibitem{Jones12}
Jones DT, Buchan DW, Cozzetto D, Pontil M (2012) {{P}{S}{I}{C}{O}{V}: precise
  structural contact prediction using sparse inverse covariance estimation on
  large multiple sequence alignments}.
\newblock {\em Bioinformatics} 28(2):184--190.

\bibitem{Dwyer13}
Dwyer RS, Ricci DP, Colwell LJ, Silhavy TJ, Wingreen NS (2013) {{P}redicting
  functionally informative mutations in {E}scherichia coli {B}am{A} using
  evolutionary covariance analysis}.
\newblock {\em Genetics} 195(2):443--455.

\bibitem{Cheng14}
Cheng RR, Morcos F, Levine H, Onuchic JN (2014) {{T}oward rationally
  redesigning bacterial two-component signaling systems using coevolutionary
  information}.
\newblock {\em Proc. Natl. Acad. Sci. U.S.A.} 111(5):E563--571.

\bibitem{Figliuzzi16}
Figliuzzi M, Jacquier H, Schug A, Tenaillon O, Weigt M (2016) {{C}oevolutionary
  {L}andscape {I}nference and the {C}ontext-{D}ependence of {M}utations in
  {B}eta-{L}actamase {T}{E}{M}-1}.
\newblock {\em Mol. Biol. Evol.} 33(1):268--280.

\bibitem{Procaccini11}
Procaccini A, Lunt B, Szurmant H, Hwa T, Weigt M (2011) {{D}issecting the
  specificity of protein-protein interaction in bacterial two-component
  signaling: orphans and crosstalks}.
\newblock {\em PLoS ONE} 6(5):e19729.

\bibitem{Baldassi14}
Baldassi C et~al. (2014) {{F}ast and accurate multivariate {G}aussian modeling
  of protein families: predicting residue contacts and protein-interaction
  partners}.
\newblock {\em PLoS ONE} 9(3):e92721.

\bibitem{Ovchinnikov14}
Ovchinnikov S, Kamisetty H, Baker D (2014) {{R}obust and accurate prediction of
  residue-residue interactions across protein interfaces using evolutionary
  information}.
\newblock {\em Elife} 3:e02030.

\bibitem{Hopf14}
Hopf TA et~al. (2014) {{S}equence co-evolution gives 3{D} contacts and
  structures of protein complexes}.
\newblock {\em Elife} 3:e03430.

\bibitem{Feinauer16}
Feinauer C, Szurmant H, Weigt M, Pagnani A (2016) {{I}nter-{P}rotein {S}equence
  {C}o-{E}volution {P}redicts {K}nown {P}hysical {I}nteractions in {B}acterial
  {R}ibosomes and the {T}rp {O}peron}.
\newblock {\em PLoS ONE} 11(2):e0149166.

\bibitem{Schneidman06}
Schneidman E, Berry MJ, Segev R, Bialek W (2006) {{W}eak pairwise correlations
  imply strongly correlated network states in a neural population}.
\newblock {\em Nature} 440(7087):1007--1012.

\bibitem{Lezon06}
Lezon TR, Banavar JR, Cieplak M, Maritan A, Fedoroff NV (2006) {{U}sing the
  principle of entropy maximization to infer genetic interaction networks from
  gene expression patterns}.
\newblock {\em Proc. Natl. Acad. Sci. U.S.A.} 103(50):19033--19038.

\bibitem{Mora10}
Mora T, Walczak AM, Bialek W, Callan CG (2010) {{M}aximum entropy models for
  antibody diversity}.
\newblock {\em Proc. Natl. Acad. Sci. U.S.A.} 107(12):5405--5410.

\bibitem{Bialek12}
Bialek W et~al. (2012) {{S}tatistical mechanics for natural flocks of birds}.
\newblock {\em Proc. Natl. Acad. Sci. U.S.A.} 109(13):4786--4791.

\bibitem{Wood12}
Wood K, Nishida S, Sontag ED, Cluzel P (2012) {{M}echanism-independent method
  for predicting response to multidrug combinations in bacteria}.
\newblock {\em Proc. Natl. Acad. Sci. U.S.A.} 109(30):12254--12259.

\bibitem{Ferguson13}
Ferguson AL et~al. (2013) {{T}ranslating {H}{I}{V} sequences into quantitative
  fitness landscapes predicts viral vulnerabilities for rational immunogen
  design}.
\newblock {\em Immunity} 38(3):606--617.

\bibitem{Mann14}
Mann JK et~al. (2014) {{T}he fitness landscape of {H}{I}{V}-1 gag: advanced
  modeling approaches and validation of model predictions by in vitro testing}.
\newblock {\em PLoS Comput. Biol.} 10(8):e1003776.

\bibitem{Casino09}
Casino P, Rubio V, Marina A (2009) {{S}tructural insight into partner
  specificity and phosphoryl transfer in two-component signal transduction}.
\newblock {\em Cell} 139(2):325--336.

\bibitem{Laub07}
Laub MT, Goulian M (2007) {{S}pecificity in two-component signal transduction
  pathways}.
\newblock {\em Annu. Rev. Genet.} 41:121--145.

\bibitem{Morcos11}
Morcos F et~al. (2011) {{D}irect-coupling analysis of residue coevolution
  captures native contacts across many protein families}.
\newblock {\em Proc. Natl. Acad. Sci. U.S.A.} 108(49):E1293--1301.

\bibitem{Jacquin16}
Jacquin H, Gilson A, Shakhnovich E, Cocco S, Monasson R (2016) Benchmarking
  inverse statistical approaches for protein structure and design with exactly
  solvable models.
\newblock {\em PLoS Comput. Biol.} 12(5):e1004889.

\bibitem{Ekeberg13}
Ekeberg M, Lovkvist C, Lan Y, Weigt M, Aurell E (2013) {{I}mproved contact
  prediction in proteins: using pseudolikelihoods to infer {P}otts models}.
\newblock {\em Phys. Rev. E} 87(1):012707.

\bibitem{AK}
Tolstoy L (1877) {\em Anna Karenina}.
\newblock Translation: R. Pevear and L. Volokhonsky (Penguin, 2001).

\bibitem{Bradde10}
Bradde S et~al. (2010) Aligning graphs and finding substructures by a cavity
  approach.
\newblock {\em EPL} 89(3).

\bibitem{Rees09}
Rees DC, Johnson E, Lewinson O (2009) {{A}{B}{C} transporters: the power to
  change}.
\newblock {\em Nat. Rev. Mol. Cell Biol.} 10(3):218--227.

\bibitem{Finn16}
Finn RD et~al. (2016) {{T}he {P}fam protein families database: towards a more
  sustainable future}.
\newblock {\em Nucleic Acids Res.} 44(D1):D279--285.

\bibitem{Barakat09}
Barakat M et~al. (2009) {{P}2{C}{S}: a two-component system resource for
  prokaryotic signal transduction research}.
\newblock {\em BMC Genomics} 10:315.

\bibitem{Ortet15}
Ortet P, Whitworth DE, Santaella C, Achouak W, Barakat M (2015) {{P}2{C}{S}:
  updates of the prokaryotic two-component systems database}.
\newblock {\em Nucleic Acids Res.} 43(Database issue):D536--541.

\bibitem{Plefka82}
Plefka T ({1982}) {Convergence condition of the TAP equation for the
  infinite-ranged Ising spin glass model}.
\newblock {\em {J. Phys. A: Math. Gen.}} {15}({6}):{1971--1978}.

\bibitem{Ekeberg14}
Ekeberg M, Hartonen T, Aurell E ({2014}) {Fast pseudolikelihood maximization
  for direct-coupling analysis of protein structure from many homologous
  amino-acid sequences}.
\newblock {\em {J. Comput. Phys.}} {276}:{341--356}.

\bibitem{Capra10}
Capra EJ et~al. (2010) {{S}ystematic dissection and trajectory-scanning
  mutagenesis of the molecular interface that ensures specificity of
  two-component signaling pathways}.
\newblock {\em PLoS Genet.} 6(11):e1001220.

\bibitem{Podgornaia13b}
Podgornaia AI, Casino P, Marina A, Laub MT (2013) {{S}tructural basis of a
  rationally rewired protein-protein interface critical to bacterial
  signaling}.
\newblock {\em Structure} 21(9):1636--1647.

\bibitem{Podgornaia15}
Podgornaia AI, Laub MT (2015) {{P}rotein evolution. {P}ervasive degeneracy and
  epistasis in a protein-protein interface}.
\newblock {\em Science} 347(6222):673--677.

\end{thebibliography}
\end{document}